\documentclass[letterpaper,twocolumn,10pt]{article}
\usepackage[available,functional, reproduced]{usenixbadges}
\usepackage{usenix2019_v3}

\usepackage{tikz}
\usepackage{amsmath}
\usepackage{xparse}
\usepackage{caption}
\usepackage{subfig}
\usepackage{setspace}
\usepackage{hyperref}
\usepackage[normalem]{ulem}
\usepackage{lipsum}
\usepackage{listings}

\usepackage{algorithmic}
\usepackage{algorithm}

\usepackage[utf8]{inputenc}
\usepackage{listings}
\usepackage{xcolor}

\usepackage{color}
\definecolor{codegreen}{rgb}{0,0.6,0}
\definecolor{codegray}{rgb}{0.5,0.5,0.5}
\definecolor{codepurple}{rgb}{0.58,0,0.82}
\definecolor{backcolour}{rgb}{0.95,0.95,0.92}
\definecolor{textblue}{rgb}{.2,.2,.7}
\definecolor{textred}{rgb}{0.54,0,0}
\definecolor{textgreen}{rgb}{0,0.43,0}
\definecolor{codered}{rgb}{201,72,12}

\usepackage[T1]{fontenc}
\usepackage[scaled=0.85]{beramono} 
\usepackage{listings}
\usepackage{multirow}
\usepackage{setspace}

\newcommand\todo[1]{\textcolor{red}{#1}}

\newcommand\subfig[1]{\textcolor{green!80!black}{#1}}

\newcommand*{\Mname}{GNNAdvisor}
\newcommand\code[1]{\texttt{\textcolor{black}{#1}}}

\newcommand\comment[1]{\textbf\textit{\textcolor{blue}{$\triangleright$~#1.}}}
\newcommand\commentone[1]{\textbf\textit{\textcolor{blue}{$\triangleright$~#1}}}
\newcommand\commenttwo[1]{\textbf\textit{\textcolor{blue}{#1.}}}

\newcommand{\yuke}[1]{{\color{blue}[Yuke: #1]}}
\algsetup{linenosize=\footnotesize}

\begin{document}

\date{}
\title{\Mname: An Adaptive and Efficient Runtime System for GNN Acceleration on GPUs}

\author{
{\rm Yuke Wang, Boyuan Feng, Gushu Li, Shuangchen Li, Lei Deng, Yuan Xie, and Yufei Ding}\\
University of California, Santa Barbara
} 

\maketitle
\pagestyle{empty}
\begin{abstract}
As the emerging trend of graph-based deep learning, Graph Neural Networks (GNNs) excel for their capability to generate high-quality node feature vectors (embeddings).
However, the existing one-size-fits-all GNN implementations are insufficient to catch up with the evolving GNN architectures, the ever-increasing graph sizes, and the diverse node embedding dimensionalities.
%
To this end, we propose \textbf{GNNAdvisor}~\footnote{Paper is accepted at OSDI'21.}, an adaptive and efficient runtime system to accelerate various GNN workloads on GPU platforms. 
First, GNNAdvisor explores and identifies several performance-relevant features from both the GNN model and the input graph, and uses them as a new driving force for GNN acceleration. 
Second, GNNAdvisor implements a novel and highly-efficient 2D workload management, tailored for GNN computation to improve GPU utilization and performance under different application settings.
Third, GNNAdvisor capitalizes on the GPU memory hierarchy for acceleration by gracefully coordinating the execution of GNNs according to the characteristics of the GPU memory structure and GNN workloads.
Furthermore, to enable automatic runtime optimization, GNNAdvisor incorporates a lightweight analytical model for an effective design parameter search.
Extensive experiments show that GNNAdvisor outperforms the state-of-the-art GNN computing frameworks, such as Deep Graph Library ($3.02\times$ faster on average) and NeuGraph (up to $4.10\times$ faster), on mainstream GNN architectures across various datasets.
\end{abstract}
\section{Introduction}
Graph Neural Networks (GNNs) 
emerge to stand on the front-line for handling many graph-based deep learning tasks (\textit{e.g.,} node embedding generation for node classification~\cite{kaspar2010graph, gibert2012graph, duran2017learning} and link prediction~\cite{chen2005link, kunegis2009learning, tylenda2009towards}).  
Compared with standard methods for graph analytics, such as random walks~\cite{grover2016node2vec, deepWalk} and graph Laplacians~\cite{luo2011cauchy, luo2009non, cheng2018deep}, GNNs highlight themselves with the interleaved two-phase execution of both graph operations (scatter-and-gather~\cite{gonzalez2012powergraph}) at the \textit{Aggregation} phase, and Neural Network (NN) operations (matrix multiplication) at the \textit{Update} phase, to achieve 
significantly higher accuracy~\cite{GCNConv, GINConv, GATConv} and better generality~\cite{SageConv}.
Yet, the state-of-the-art GNN frameworks~\cite{wang2019dgl, pyG, ma2019neugraph, wang2016gunrock}, which follow a one-size-fits-all implementation scheme, often suffer from poor performance when handling more complicated GNN architectures (\textit{i.e.,} more layers and higher hidden dimensionality in each layer) 
and diverse graph datasets.
\begin{figure*} [t] \small
    \centering
    \includegraphics[width=0.95\textwidth]{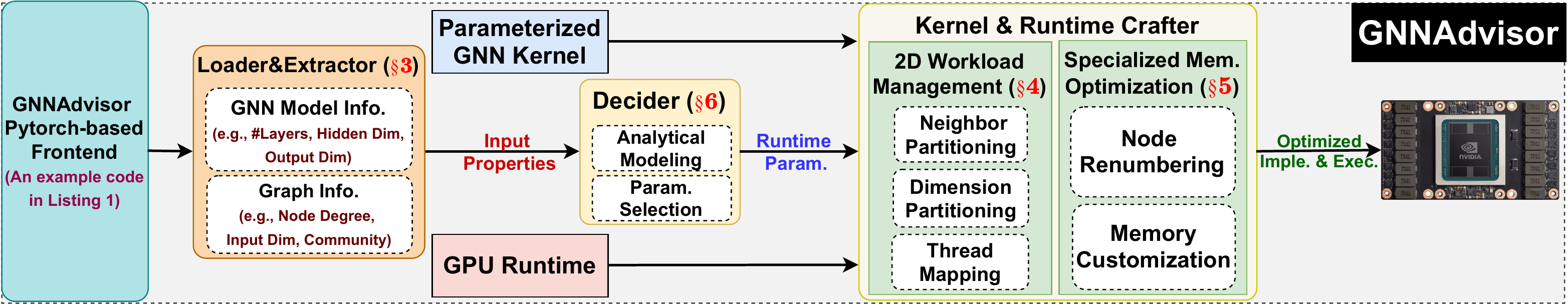}
    \vspace{-5pt}
    \caption{Overview of \Mname.}
    \label{fig: Overview}
    \vspace{-10pt}
\end{figure*}

Specifically, previous work that supports both GNN training and inference can be classified into two categories.
The first type~\cite{wang2016gunrock, ma2019neugraph} is built on popular graph processing systems and is combined with NN operations.
The second type~\cite{pyG, wang2019dgl}, in contrast, starts with deep learning frameworks and is extended to support vector-based graph operations.
However, these existing solutions are still preliminary and inevitably fall short in the following three major aspects, even on common computing platforms such as GPUs. 

\textbf{Failing to leverage GNN input information.} 
GNN models demonstrate great diversity in terms of layer sequences, types of aggregation methods, and the dimension size of node embeddings. 
These profoundly impact the effectiveness of various system optimization choices.  
The diversity of input graphs further complicates the problem.
Unfortunately, current GNN frameworks~\cite{pyG,wang2019dgl,ma2019neugraph} follow a one-size-fits-all optimization scheme and fail to craft an optimization strategy that maximizes efficiency for a particular GNN application's settings.
%
Some classical graph systems~\cite{balaji2018graph,rabbit-order} have exploited input characteristics to facilitate more efficient optimizations, but they only focus on simple graph algorithms like PageRank~\cite{page1999pagerank} while having no support for GNN models.  

\textbf{Optimizations not tailored to GNN.}
While the update phase in GNNs involves NN operations that are dense in computation and regular in memory access, the aggregation phase is usually sparse in computation and highly irregular in memory access. 
Without dedicated optimization, it will inevitably become the performance bottleneck. 
Existing GNN frameworks~\cite{pyG,wang2019dgl,ma2019neugraph} simply extend the optimization schemes from classical graph systems~\cite{khorasani2014cusha, wang2016gunrock}, and do not address the difference between GNN and graph processing. 
For example, each node is associated with an embedding attribute in GNNs while each node only has a single scalar attribute in traditional graph processing.
Such difference invokes novel design principles for GNNs towards more aggressive parallelism and more efficient memory optimization. 

\begin{minipage}[H]{0.45\textwidth}
\centering
\vspace{-5pt}
\begin{lstlisting}[caption=Example of a 2-layer GCN in GNNAdvisor.]
import GNNAdvisor as GNNA
import torch
# import other packages ...

# Create a GCN class.
class GCN(torch.nn.Module):
    def __init__(self, inDim, hiDim, outDim, nLayers):
        self.layers = torch.nn.ModuleList()
        self.layers.append(GNNA.GCNConv(inDim, hiDim))
        for i in range(nLayers - 2):
            layer = GNNA.GCNConv(hiDim, hiDim)
            self.layers.append(layer)
        self.layers.append(GNNA.GCNConv(hiDim, outDim))
        self.softmax = torch.nn.Softmax()
        
    def forward(self, X, graph, param):
        for i in range(len(self.layers)):
            X = self.layers[i](X, graph, param)
            X = self.ReLU(X)
        X = self.softmax(X)
        return X

# Define a two-layer GCN model.
model = GCN(inDim=100, hiDim=16, outDim=10, nLayers=2)

# Loading graph and extracting input propertities.
graphObj, inputInfo = GNNA.LoaderExtractor(graphFile, 
                                            model)
# Set runtime parameters automatically.
X, graph, param = GNNA.Decider(graphObj, inputInfo)

# Run model.
predict_y = model(X, graph, param)

# Compute loss and accuracy.
# Gradient backpropagation for training.
\end{lstlisting} 
\label{code: GCN code example.}
\end{minipage}

\textbf{Poor runtime support for input adaptability.}
%
 Prior GNN frameworks~\cite{pyG,wang2019dgl,ma2019neugraph} rely on a Python-based high-level programming interface for ease of user implementation. 
These frameworks employ static optimizations through a compiler or manually-optimized libraries. Nevertheless, some critical performance-related information for a GNN is only available at runtime (\textit{e.g.}, node degree and embedding size). 
Without adaptable designs that can leverage such runtime information, we would easily suffer from an inferior performance because of the largely under-utilized the GPU computing resources and inefficient irregular memory access.
This limitation motivates the need for runtime environments with flexible designs to handle a wide spectrum of inputs effectively.

To this end, we propose, \Mname, an adaptive and efficient runtime system for GNN acceleration on GPUs. 
\Mname~leverages Pytorch~\cite{pytorch} as the front-end to improve programmability and ease user implementation. 
We show a representative 2-layer Graph Convolutional Network (GCN)~\cite{GCNConv} in \Mname~at Listing~\ref{code: GCN code example.}. 
At the low level, GNNAdvisor is built with C++/CUDA and integrated with Pytorch framework by using Pytorch Wrapper. 
It can be viewed as a new type of Pytorch operator with a set of kernel optimizations and runtime support. It can work seamlessly with existing operators from the Pytorch Framework. 
Data is loaded with the data loader written in Pytorch and passed as a Tensor to GNNAdvisor for computation on GPUs. Once the GNNAdvisor completes its computation at the GPU, it will pass the data Tensor back to the original Pytorch framework for further processing.
As detailed in Figure~\ref{fig: Overview}, \Mname~consists of several key components to facilitate the GNN optimization and execution on GPUs. 
First, \Mname~introduces an input \textbf{\code{Loader\&Extractor}} to exploit the input-level information that can guide our system-level optimizations.
Second, \Mname~incorporates a \textbf{\code{Decider}} consisting of analytical modeling for automatic runtime parameter selection to reduce manual effort in design optimization, and a lightweight node renumbering routine to improve graph structural locality.
Third, \Mname~integrates a \textbf{\code{Kernel\&Runtime Crafter}} to customize our parameterized GNN kernel and CUDA runtime, which consists of an effective 2D workload management (considering both the number of neighbor nodes and the node embedding dimensionality) and a set of GNN-specialized memory optimizations.
%

Note that in this project, we mainly focus on the setting of single-GPU GNN computing, which is today's most popular design adopted as the key component in many state-of-the-art frameworks, such as DGL~\cite{wang2019dgl} and PyG~\cite{pyG}. Single-GPU GNN computing is desirable for two reasons:
First, many GNN applications with small to medium size graphs (\textit{e.g.}, molecule structure) can easily fit the memory of a single GPU. 
Second, in the case of large-size graphs that can only be handled by out-of-GPU-core and multi-GPU processing, numerous well-studied graph partition strategies (\textit{e.g.}, METIS~\cite{METIS}) can cut the giant graphs into small-size subgraphs to make them suitable for a single GPU. Therefore, the optimization of both the out-of-GPU-core (\textit{e.g.}, GPU streaming processing) and multi-GPU GNN computation still largely demands performance improvements on a single GPU.
Moreover, while our paper focuses on GNNs, our proposed methodology can be applied to optimize various types of irregular workload (\textit{e.g.}, social network analysis) targeting GPUs as well.

Overall, we make the following contributions:
\begin{itemize}
    \item  We are the first to explore GNN input properties (\textcolor{red!70!black}{$\S3$}) (\textit{e.g.}, GNN model architectures and input graphs), and give an in-depth analysis of their importance in guiding system optimizations for GPU-based GNN computing.
    \item We propose a set of GNN-tailored system optimizations with parameterization, including a novel 2D workload management (\textcolor{red!70!black}{$\S4$}) and specialized memory customization (\textcolor{red!70!black}{$\S5$}) on GPUs. We incorporate the analytical modeling and parameter auto-selection (\textcolor{red!70!black}{$\S6$}) to ease the design space exploration.
    \item Comprehensive experiments demonstrate the strength of \Mname~over state-of-the-art GNN execution frameworks, such as Deep Graph Library (average $3.02\times$) and NeuGraph (average $4.36\times$), on mainstream GNN architectures across various datasets.
\end{itemize}

\section{Background and Related Work} 
\label{sect: Related Works}
 In this section, we introduce the basics of GNNs and two major types of GNN computing frameworks: \textit{GPU-based graph systems} and \textit{deep learning frameworks}. 

\vspace{-5pt}
\subsection{Graph Neural Networks}
\label{sect: Graph Neural Networks}
\begin{figure} [h] \small
    \centering
    \includegraphics[width=0.8\columnwidth]{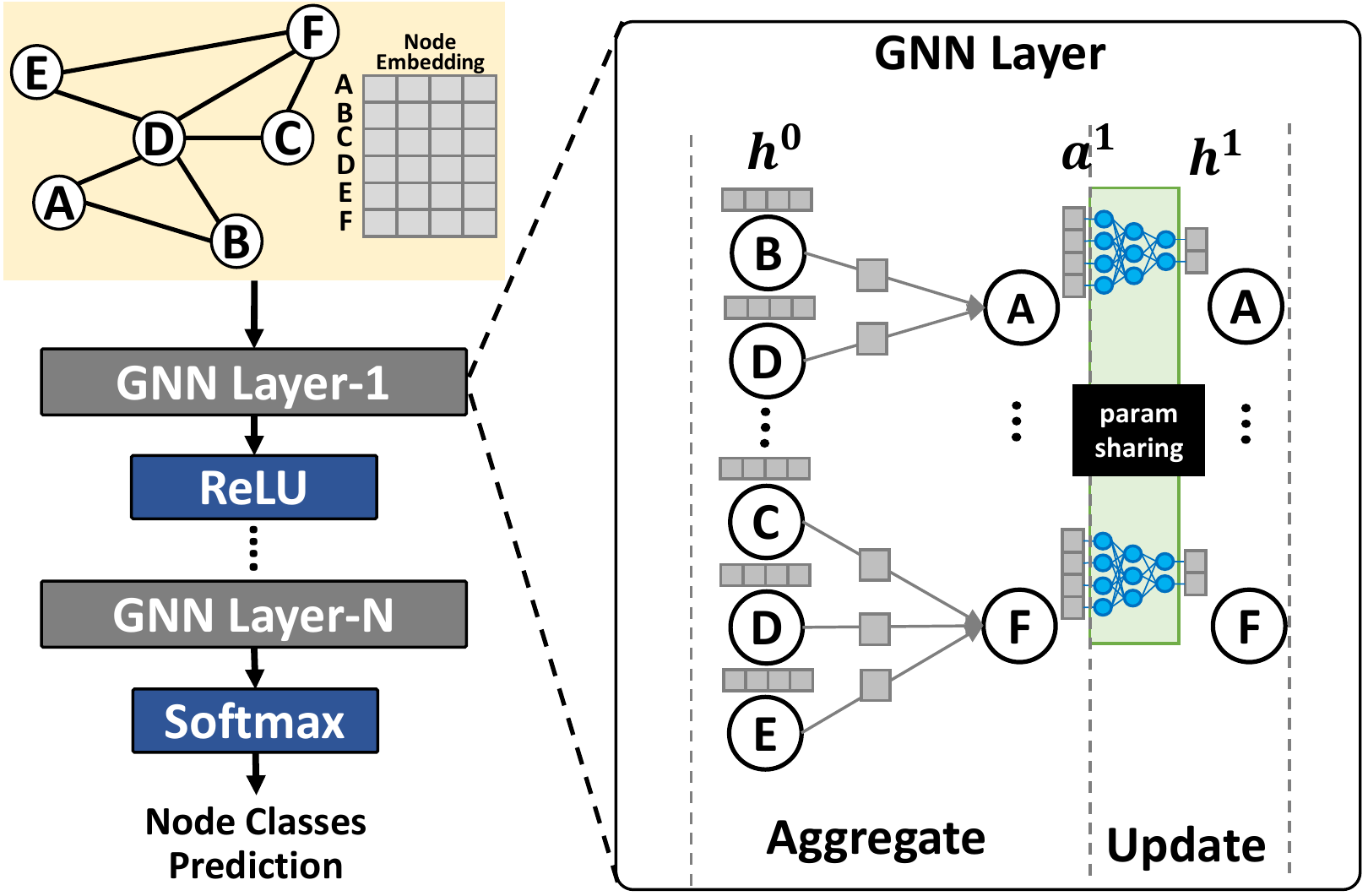}
    \vspace{-5pt}
    \caption{GNN General Computation Flow.}
    \label{fig: GNN Computation Flow}
     \vspace{-5pt}
\end{figure}

Figure~\ref{fig: GNN Computation Flow} visualizes the computation flow of GNNs in one iteration. 
GNNs compute the node feature vector (embedding) for node $v$ at layer $k+1$ based on the embedding information at layer $k$ ($k \geq 0$), as shown in Equation~\ref{eq: GNN},
\begin{gather} \small \label{eq: GNN}
 \begin{aligned} 
   a_{v}^{(k+1)}  &= \mathbf{Aggregate}^{(k+1)}({h_{u}^{(k)}|u\in \mathbf{N}(v)\cup h_v^{(k)}}) \\
   h_{v}^{(k+1)}  &= \mathbf{Update}^{(k+1)}(a_{v}^{(k+1)})
\end{aligned}   
\end{gather}
where $h_{v}^{(k)}$ is the embedding vector for node $v$ at layer $k$; $h_v^{(0)}$ is computed from the task-specific features of a vertex (e.g., the text associated with the vertex, or some scalar properties of the entity that the vertex represents) via some initial embedding mapping that is used only for this ingest of symbolic values into the embedding space; $a_{v}^{(k+1)}$ is the aggregation results through collecting neighbors' information (\textit{e.g.}, node embeddings); $\mathbf{N}(v)$ is the neighbor set of node $v$.
The aggregation method and the order of aggregation and update could vary across different GNNs. 
Some methods~\cite{GCNConv, SageConv} just rely on the neighboring nodes while others~\cite{GATConv} also 
leverage edge properties, by combining the dot product of the end-point nodes of each edge, along with any edge features (edge type and other attributes).
The update function is generally composed of standard NN operations, such as a single fully connected layer or a multi-layer perceptron (MLP) in the form of $w\cdot a_{v}^{(k+1)} + b$, where $w$ and $b$ are the learnable weight and bias parameters, respectively.
The common choices for node embedding dimensions are 16, 64, and 128, and the embedding dimension may change across different layers. 

After passing through several iterations of aggregation and update (\textit{i.e.}, several GNN layers), we will get the output embedding of each node, which can usually be used for various downstream graph-based deep learning tasks, such as node classification~\cite{kaspar2010graph, gibert2012graph, duran2017learning} and link prediction~\cite{chen2005link, kunegis2009learning, tylenda2009towards}. 
Note that the initial node embedding for GNN's input layer may come with the original graph dataset or can be generated by a set of graph embedding algorithms, such as~\cite{grover2016node2vec, transE, duvenaud2015convolutional}, which is not included in the computation of GNNs models (generating the hidden and output node embeddings).

\subsection{Graph Processing Systems} 
\label{sect: Graph-based System}
Numerous graph processing systems~\cite{khorasani2014cusha, Tigr, wang2016gunrock, liu2015enterprise, liu2019simd} have been proposed to accelerate traditional graph algorithms. 
The major commonalities of these systems include the vertex/node-centric programming abstraction, edge-centric processing, and system optimizations to reduce the computation irregularity (\textit{e.g.}, workload imbalance) and memory access irregularity (\textit{e.g.}, non-coalesced global memory access).
However, extending these graph processing systems to support GNN computing meets with substantial challenges.

First, the common algorithm optimizations in graph processing may not benefit GNNs.
For example, graph traversal algorithms, such as Breadth-first Search, rely on iterative computing on node frontiers (active neighbors). 
Therefore, a set of frontier-based optimizations, such push-pull traversal~\cite{liu2015enterprise,liu2019simd}, and frontier filtering~\cite{liu2015enterprise,liu2019simd,wang2016gunrock}, have been extensively studied. 
However, GNNs consistently maintain fixed-sized frontiers (all neighbors) of each node across iterations. 

Second, the system optimization techniques for graph processing would benefit GNNs only after careful adaption and calibration.
For example, node/edge-centric
processing~\cite{wang2016gunrock,liu2019simd} 
and shard-based graph representation~\cite{khorasani2014cusha} 
are tailored for processing nodes/edges represented with a single scalar attribute. 
In GNNs, there's another dimension for data parallelism, namely the embedding dimension, which tends to be large
Therefore, previous design trade-offs between the coarse-grained node-level parallelism and node-value locality should be further extended to balance dimension-wise parallelism and node-embedding locality at a finer granularity.   

Third, some essential functionalities of GNN computing are missing in graph systems. For example, the node update based on NN computing for both the forward value propagation and the complicated backward gradient propagation is not available in graph systems~\cite{khorasani2014cusha, Tigr,wang2016gunrock,liu2015enterprise,liu2019simd,kyrola2012graphchi, x-stream}.
In contrast, Pytorch~\cite{pytorch} and Tensorflow~\cite{tensorflow2015} feature an analytic differentiation function for automatic gradient computations on various deep learning model architectures and functions. 
Therefore, extending the graph-processing system to support GNN computing requires non-trivial efforts, and thus we develop \Mname~on top of a deep learning framework. 

\vspace{-5pt}
\subsection{Deep Learning Frameworks}
\vspace{-5pt}
Various NN frameworks have been proposed, such as Tensorflow~\cite{tensorflow2015}, and Pytorch~\cite{pytorch}. 
These frameworks provide the end-to-end training and inference support for traditional deep-learning models with various NN operators, such as linear and convolutional operators.
These operators are highly optimized for Euclidean data (\textit{e.g.}, image) but lack support for non-Euclidean data (\textit{e.g.}, graph) in GNNs. 
Extending NN frameworks to support GNN that takes the highly-irregular graphs as the input is facing several challenges. 

First, NN-extended GNN computing platforms~\cite{pyG,wang2019dgl} focus on programmability and generality for different GNN models but lack efficient backend support to achieve high performance. 
For example, Pytorch-Geometric (PyG)~\cite{pyG} uses the torch-scatter~\cite{torch-scatter} library implemented with CUDA as its major building block of graph aggregation operations. 
The torch-scatter implementation scales poorly when encountering large sparse graphs with high-dimensional node embedding because its kernel design essentially borrows the design principles of graph-processing systems by using excessive high-overhead atomic operations to support node embedding propagation. 
A similar scalability problem is also observed in Deep Graph Library (DGL)~\cite{wang2019dgl}, which incorporates an off-the-shelf Sparse-Matrix Multiplication (SpMM) (\textit{e.g.}, \textit{csrmm2} in cuSparse~\cite{cusparse}) for simple sum-reduced aggregation~\cite{GCNConv,SageConv} and leverages its own CUDA kernel for more complex aggregation scheme with edge attributes~\cite{GINConv,GATConv}. 

Second, major computation kernels~\cite{pyG, wang2019dgl} are hard-coded without design flexibility, which is essential to handle diverse application settings with different input graph sizes and node embedding dimensionality. 
From the high-level interface, users are only allowed to define the way of composing these kernels externally.
Users are not allowed to customize kernels internally based on the known characteristics of GNN model architectures, GPU hardware, and graph properties.

\section{Input Analysis of GNN Applications} 
\label{sect: Input Extractor}
In this section, we argue that the GNN input information can guide the system optimization, based on our key observation that different GNN application settings would favor different optimization choices. 
We introduce two types of GNN input information and discuss their potential performance benefits and extraction methods.

\subsection{GNN Model Information}
\label{sect: GNN Architecture}
While the GNN update phase follows a relatively fixed computing pattern, the GNN aggregation phase shows high diversity.
The mainstream aggregation methods of GNNs can be categorized into two types: 

The first type is aggregation (\textit{e.g.}, \textit{sum}, and \textit{min}) with only the embeddings of neighbor nodes, as in Graph Convolutional Network (GCN)~\cite{GCNConv}.
For GNNs with this type of aggregation, the common design practice is to reduce the node embedding dimensionality during the update phase
(\textit{i.e.}, multiplying the node embedding matrix with the weight matrix)~\cite{GCNConv, pyG, wang2019dgl}
before the aggregation 
(gather information from neighbor node embedding) 
at each GNN layer, thereby, largely reducing the data movements during the aggregation. In this case, improving memory locality would be more beneficial, in that more node embeddings can be cached in fast memory (\textit{e.g.}, L1 cache of GPUs) to exploit performance benefits. 

The second type is aggregation with special edge features (\textit{e.g.}, weights, and edge vectors that are computed by combining source and target nodes) applied to each neighbor node, as in Graph Isomorphism Network (GIN)~\cite{GINConv}.
This type of GNN must work on large full-dimensional node embeddings to compute the special edge features at the node aggregation.
In this case, the fast memory (\textit{e.g.}, shared memory of GPU Stream-Multiprocessors) is not large enough to exploit memory locality.
However, improving computation parallelization (\textit{e.g.}, workload partitioning along the embedding dimension) would be more helpful, considering that workloads can be shared among more concurrent threads for improving overall throughput. 

We illustrate this aggregation-type difference with the mathematical equations for GCN and GIN. With GCN, the output embedding $\mathbf{X}$ is computed as follows:
\begin{equation} \small
    \mathbf{X}^{\prime} = \mathbf{\hat{D}}^{-1/2} \mathbf{\hat{A}}
    \mathbf{\hat{D}}^{-1/2} \mathbf{X} \mathbf{W},
\end{equation}
{where $\mathbf{\hat{D}}$ is the diagonal node degree matrix; $\mathbf{W}$ is the weight matrix; $\mathbf{\hat{A}}$ is the graph adjacency matrix.} For GIN, the output embedding $\mathbf{X}$ for each layer is computed as follows:
\begin{equation}
    \mathbf{x}^{\prime}_i = h \left( (1 + \epsilon) \cdot
            \mathbf{x}_i + \sum_{j \in \mathcal{N}(i)} \mathbf{x}_j \right)    
\end{equation}
where $h$ denotes a neural network, \textit{e.g.}, an MLP, which maps node features $x$ with input embedding dimension and output embedding dimension; $\epsilon$ is a configurable/trainable parameter depending on the users' demands or application settings;  $\mathcal{N}(i)$ denotes the neighbor IDs of the node $i$.

Assume we have GCN and GIN with hidden dimension 16, and the input dataset has a node embedding dimension of 128. In the case of GCN, we will first do node update (GEMM\footnote{General Matrix-Matrix Multiplication.}-based linear transformation) of the node embedding, thus, at the aggregation, we only need to do aggregation on nodes with hidden dimension 16. In the GIN case, we have to do neighbor aggregation on nodes with 128 dimensions then do node update to linearly transform node embedding from 128 to 16 dimensions. Such an aggregation difference would also lead to different optimization strategies, where GCN would prefer more memory optimization on the small node embeddings while GIN would prefer more computing parallelism on the large node embeddings.

To conclude, the type of aggregation in GNNs should be considered for system-level optimization and it can be obtained by \Mname's built-in parser of GNN model proprieties.

\subsection{Graph Information} 
\label{sect: Graph Information}
\textbf{Node Degree \& Embedding Dimensionality: } 
Real-world graphs generally follow the power-law distribution~\cite{graph-power-law} of node degrees. 
Such distribution already causes workload imbalance 
in traditional graph processing systems~\cite{liu2015enterprise, Mizan, han2014experimental}. 
In GNN aggregation, such workload imbalance 
would be exacerbated due to the higher dimensionality of the node embeddings if we perform node-centric workload partitioning. 
Moreover, node embedding would invalidate some cache-based optimizations that are originally applied to graph processing systems, since caches are usually small in size and insufficient to hold enough nodes with their embeddings. 
For example, in the graph processing scenarios with a scalar attribute for each node, we can improve performance by putting $16\times10^3$ nodes on the $64$KB L1 cache of each GPU thread block.
However, in typical GNNs with a 64-dimension embedding for each node, we can only fit 256 nodes on each GPU block's cache.


With node degree and embedding dimensionality information, new optimization opportunities for GNNs may appear because
we can estimate the node's workload and its concrete composition based on such input information. 
If the workload size is dominated by the number of node neighbors (\textit{e.g.}, large node degree), we may customize the design that could concurrently process more neighbors to increase the computing parallelism among neighbors. 
On the other hand, if the workload size is dominated by node embedding size (\textit{e.g.}, high-dimensional node embedding), we may consider boosting the computing parallelism along the node embedding dimension. 
Note that the node degree and embedding dimension information can be extracted based on the loaded graph structure and node embedding vectors.
\Mname~manages the GNN workload based on such information (Section~\ref{sect: 2D Workload Management}).

\begin{figure} [t] \small
    \centering
    \includegraphics[width=0.98\columnwidth]{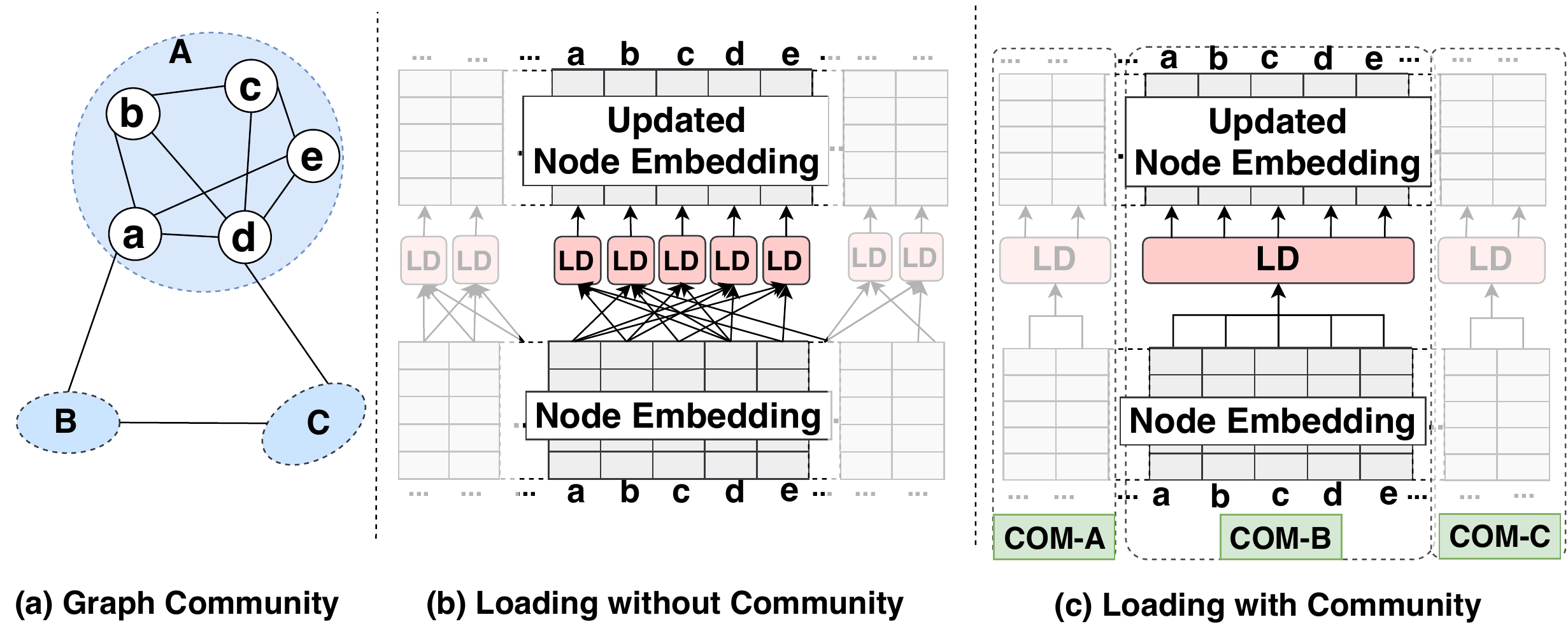}
    \vspace{-10pt}
    \caption{Graph community and its potential benefits. Note that ``LD'': loading operation. ``COM'': community.}
    \label{fig: Graph community}
    \vspace{-0pt}
\end{figure}

\textbf{Graph Community: } 
Graph community~\cite{fortunato2010community, lancichinetti2008benchmark, newman2013spectral} is one key feature of real-world graphs, which describes that a small group of nodes tend to hold ``strong'' intra-group connections (many edges) while maintaining ``weak'' connections (fewer edges) with the remaining part of the graph.
A motivating example of GNN optimization with graph community structure is shown in Figure~\ref{fig: Graph community}\subfig{a}.
Existing node-centric aggregation employed by many graph processing systems~\cite{wang2016gunrock, khorasani2014cusha} is shown in Figure~\ref{fig: Graph community}\subfig{b}, where each node will first load its neighbors and then do aggregation independently. 
This strategy can achieve great computation parallelism when each neighbor has a lightweight scalar attribute. 
In this case, the benefit of loading parallelization would offset the downside of duplicate loading of some shared neighbors.
However, in GNN computing where node embedding size is large, this node-centric loading would trigger significant unnecessary memory access since the cost of duplicate neighbor loading is now dominant and not offset by per-node parallelism
For example, aggregation of node $a$, $b$, $c$, $d$, and $e$ would load the embeddings of 15 nodes in total and most of these loads are repeated (both node $a$ and $b$ load the same node $d$ during the aggregation). 
Such loading redundancy is exacerbated with the increase of embedding dimensionality.  
On the other side, by considering the community structure of real-world graphs, unnecessary data loading for these ``common'' neighbors can be well reduced (Figure~\ref{fig: Graph community}\subfig{c}), where aggregation only requires loads of 5 distinct nodes. 

This idea sounds promising, but the effort to realize its benefits on GPUs is non-trivial. 
Existing approaches~\cite{hendrickson2000graph, newman2013spectral} of exploiting the graph communities mainly target CPU platforms with a limited number of parallelized threads and MB-level cache sizes for each thread. 
Their major goal is to exploit the data locality for every single thread. 
GPUs, on the other side, are equipped with a massive number of parallel threads and KB-level cache sizes per thread.
Therefore, the key to exploiting graph community on GPUs is to effectively exploit the data locality among threads by leveraging the L1 cache. 
Specifically, we need first capture the communities of a graph and then map such locality from input level (node-ID adjacency) to underlying GPU kernels (thread/warp/block-ID adjacency). 
The major hardware-level insight is that threads close in their IDs are more likely to share memory and computing resources, thus, improving the data spatial and temporal locality. 
\Mname~handles all these details through community-aware node renumbering and GNN-specialized memory optimizations (Section~\ref{sect: Specialized Memory Optimization}).

\section{2D Workload Management} 
\label{sect: 2D Workload Management}
GNNs employ a unique space in graph computations, due to the representation of each node by a high-dimensional feature vector (the embedding).
GNN workloads grow in two major dimensions:
\textit{the number of neighbors} and \textit{the size of the embedding dimension}. 
\Mname~incorporates an input-driven parameterized 2D workload management tailored for GNNs, including three techniques: \textit{coarse-grained neighbor partitioning}, \textit{fine-grained dimension partitioning}, and \textit{warp-based thread alignment}.

\subsection{Coarse-grained Neighbor Partitioning} 
Coarse-grained neighbor partitioning is a novel workload balance technique tailored to GNN computing on GPUs. It aims to tackle the challenge of \textit{inter-node workload imbalance} and \textit{redundant atomic operations}. 
\begin{figure} [t] \small
    \centering
    \includegraphics[width=\columnwidth]{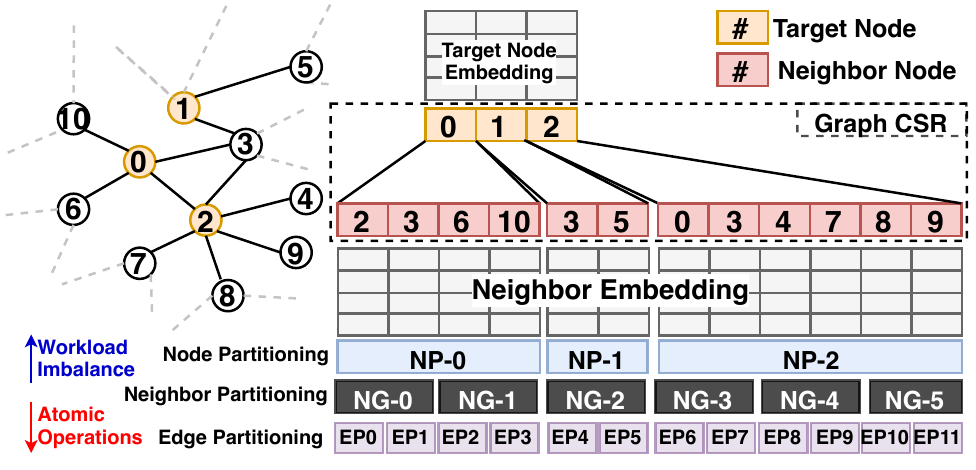}
    \caption{Neighbor Partitioning. Note that ``NP'': Node Partitioning; ``EP'': Edge Partitioning; ``NG'': Neighbor Group.}
    \label{fig: Group-based aggregration}
\end{figure}

Specifically, based on the loaded graph compressed-sparse row (CSR) representation, our coarse-grained neighbor partitioning will first break down the neighbors of a node into a set of equal-sized neighbor groups, and treat the aggregation workload of each neighbor group (NG) as the basic workload unit for scheduling. 
Figure~\ref{fig: Group-based aggregration} exemplifies an undirected graph and its corresponding neighbor partitioning result.
The neighbors of Node-0 are divided into two neighbor groups (NG-0 and NG-1) with a pre-determined group size of 2. Neighbors (Node-3 and Node-5) of Node-1 are covered by NG-2, while the neighbors of Node-2 are spread among NG-\{3,4,5\}. 
To support the neighbor group, we introduce two components, the neighbor-partitioning module and the neighbor-partitioning graph store. 
The former is a lightweight module built on top of the graph loader by partitioning the graph CSR into equal-size groups. Note that each neighbor group only covers the neighbors of one target node for ease of scheduling and synchronization.
The neighbor-partitioning graph store maintains the tuple-based meta-data of each neighbor group, including its IDs, starting and ending position of its neighbor nodes in the CSR representation, and the source node. For example, the meta-data of NG-2 will be stored as (2, 1, (4, 6)), where 2 is the neighbor-group ID, 1 is the target node ID, (4, 6) is the index range of the neighbor nodes in CSR.

The benefits of applying the aggregation based on partitioning neighbors are three-fold: 
1) compared with the more coarse-grained aggregation based on node/vertex-centric partitioning~\cite{khorasani2014cusha},
neighbor partitioning can largely mitigate the size irregularity of the workload units, which would improve GPU occupancy and throughput performance;
2) compared with the more fine-grained edge-centric partitioning 
(used by existing GNN frameworks, such as PyG~\cite{pyG}, for batching and tensorization, and graph processing systems~\cite{liu2019simd ,wang2016gunrock} for massive computing parallelization), 
the neighbor-partitioning solution can avoid the overheads of managing many tiny workload units that might hurt the performance in many ways, such as scheduling overheads and the excessive amount of synchronizations; 
3) it introduces a performance-related parameter, \textbf{neighbor-group} size ($ngs$), which is used for design parameterization and performance tuning. Neighbor partitioning works at a coarse granularity of individual neighbor nodes. It can largely mitigate the workload imbalance problem for low-dimension settings. 
For high-dimensional node embeddings, we employ a fine-grained dimension partitioning discussed in the next subsection to further distribute workloads of each neighbor group to threads.
Note that when the number of neighbors is not divisible by the neighbor group size, it will raise neighbor-group imbalance. 
Such irregularity can be amortized by setting the neighbor-group size to a small number (\textit{e.g.}, 3). 


\subsection{Fine-grained Dimension Partitioning}
\label{sect: Fine-grained Dimension Partitioning}
\begin{figure} [t] \small
    \centering
    \includegraphics[width=\columnwidth]{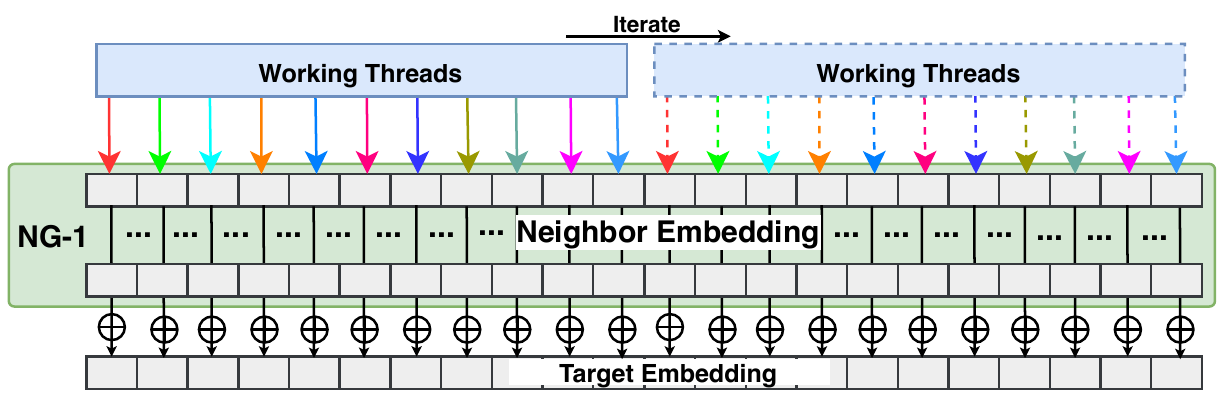}
    \caption{Dimension Partitioning. $\bigoplus$: Accumulated add.}
    \vspace{-15pt}
    \label{fig: Dimension-based Workload Sharing.}
\end{figure}

GNN distinguishes itself from traditional graph algorithms in its computation on the node embedding. 
To explore the potential acceleration parallelism along this dimension, we leverage a fine-grained dimension partitioning to further distribute the workloads of a neighbor group along the embedding dimension to improve aggregation performance. 
As shown in Figure~\ref{fig: Dimension-based Workload Sharing.}, the original neighbor-group workloads are evenly distributed to 11 consecutive threads, where each thread manages the aggregation along one dimension independently (\textit{i.e.}, accumulation of all neighbor node embeddings towards the target node embedding). 
If the dimension size is larger than the number of working threads, more iterations would be required to finish the aggregation. 

There are two major reasons for using dimension partitioning. 
First, it can accommodate a more diverse range of embedding dimension sizes. 
We can either increase the number of concurrent dimension workers or enable more iterations to handle the dimension variation flexibly. 
This is essential for modern GNNs with increasingly complicated model structures and different sizes of embedding dimension.
Second, it introduces another performance-related parameter -- the number of working threads (\textbf{dimension-worker} ($dw$)) for design customization. The value of this parameter can help to balance the thread-level parallelism and the single thread efficiency (\textit{i.e.}, computation workload per thread). 

\subsection{Warp-based Thread Alignment} 
\begin{figure} [t] \small
    \centering
    \includegraphics[width=\columnwidth]{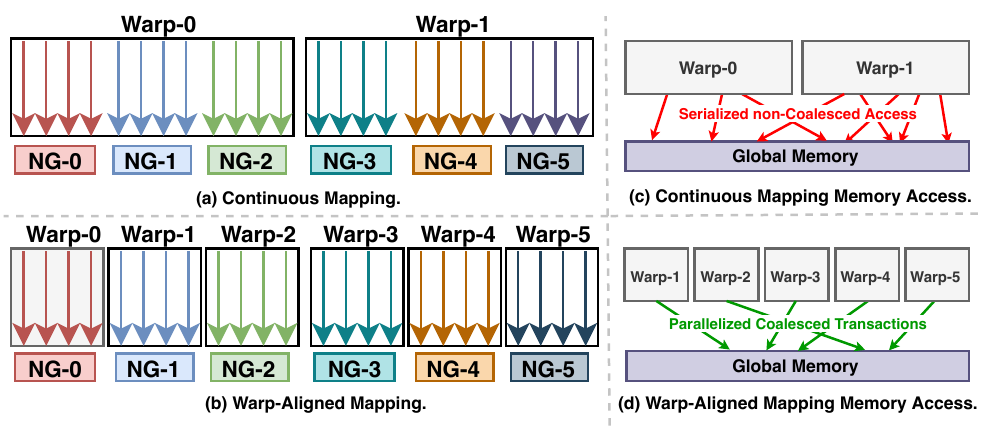}
    \vspace{-15pt}
    \caption{Warp-based Thread Alignment. }
    \vspace{-15pt}
    \label{fig: warp-based Thread Alignment}
\end{figure}

While the above two techniques answer how we balance GNN workloads logically, 
how to map these workloads to underlying GPU hardware for efficient execution is still unresolved. 
One straightforward solution is to assign consecutive threads to concurrently process workloads from different neighbor groups (Figure~\ref{fig: warp-based Thread Alignment}\subfig{a}).
However, different behaviors (\textit{e.g.}, data manipulation and memory access operations) among these threads would result in thread divergence and GPU underutilization. 
Threads from the same warp proceed in a single-instruction-multiple-thread (SIMT) fashion and the warp scheduler can only serve one type of instruction per cycle. Therefore, different threads have to wait for their turn for execution until the Stream-Multiprocessor (SM) warp scheduler issues their corresponding instructions.

To tackle this challenge, we introduce a warp-aligned thread mapping in coordination with our neighbor and dimension partitioning 
to systematically capitalize on the performance benefits of balanced workloads.
As shown in Figure~\ref{fig: warp-based Thread Alignment}\subfig{b}, each warp will independently manage the aggregation workload from one neighbor group. 
Therefore, the execution of different neighbor groups (\textit{e.g.}, NG-0 to NG-5) can be well parallelized without inducing warp divergence. 
There are several benefits in employing warp-based thread alignment.
First, inter-thread synchronization (\textit{e.g.}, atomic operations) can be minimized. Threads of the same warp are working on different dimensions of the same neighbor group, thus no conflicts occur for either global or shared memory accesses by threads from the same warp.

Second, the workload of a single warp is reduced and different warps will process more balanced workloads. 
Therefore, more small warps can be managed flexibly by SM warp schedulers to improve overall parallelism. 
Considering the
unavoidable global memory access of each warp during aggregation, increasing the number of warps can improve SM occupancy to hide latency.
Third, memory access can be coalesced. 
Threads with consecutive IDs from the same warp will access continuous memory addresses in global memory for node embeddings. 
Therefore, compared with continuous thread mapping (Figure~\ref{fig: warp-based Thread Alignment}\subfig{c}), warp-aligned thread mapping can merge memory requests from the same warp into one global memory transaction (Figure~\ref{fig: warp-based Thread Alignment}\subfig{d}).
\section{Specialized Memory Optimization} 
\label{sect: Specialized Memory Optimization}
To further exploit the benefits of 2D workload, we introduce GNN-specialized memory optimizations, \textit{community-aware node renumbering} and \textit{warp-aware memory customization}.

\subsection{Community-aware Node Renumbering} 
\label{sect: Node Renumbering}
To explore the performance benefits of graph community (Section~\ref{sect: Graph Information}), we incorporate lightweight node renumbering by reordering node IDs to improve the temporal/spatial locality during GNN aggregation without compromising output correctness.
The key idea is that the proximity of node IDs would project to the adjacency of computing units on GPU where they get processed. 
In \Mname, our 2D workload management assigns neighbor groups of a node to consecutive warps based on their node ID. 
If two nodes are assigned with consecutive IDs, 
their corresponding neighbor groups (warps) would be close to each other in their warp IDs as well. 
Thus, they are more likely to be scheduled closely on the same GPU SM with a shared L1 cache to improve the data locality on loaded common neighbors. To apply node renumbering effectively, two key questions must be addressed.

\begin{figure} [t] \small
    \centering
    \includegraphics[width=0.9\columnwidth]{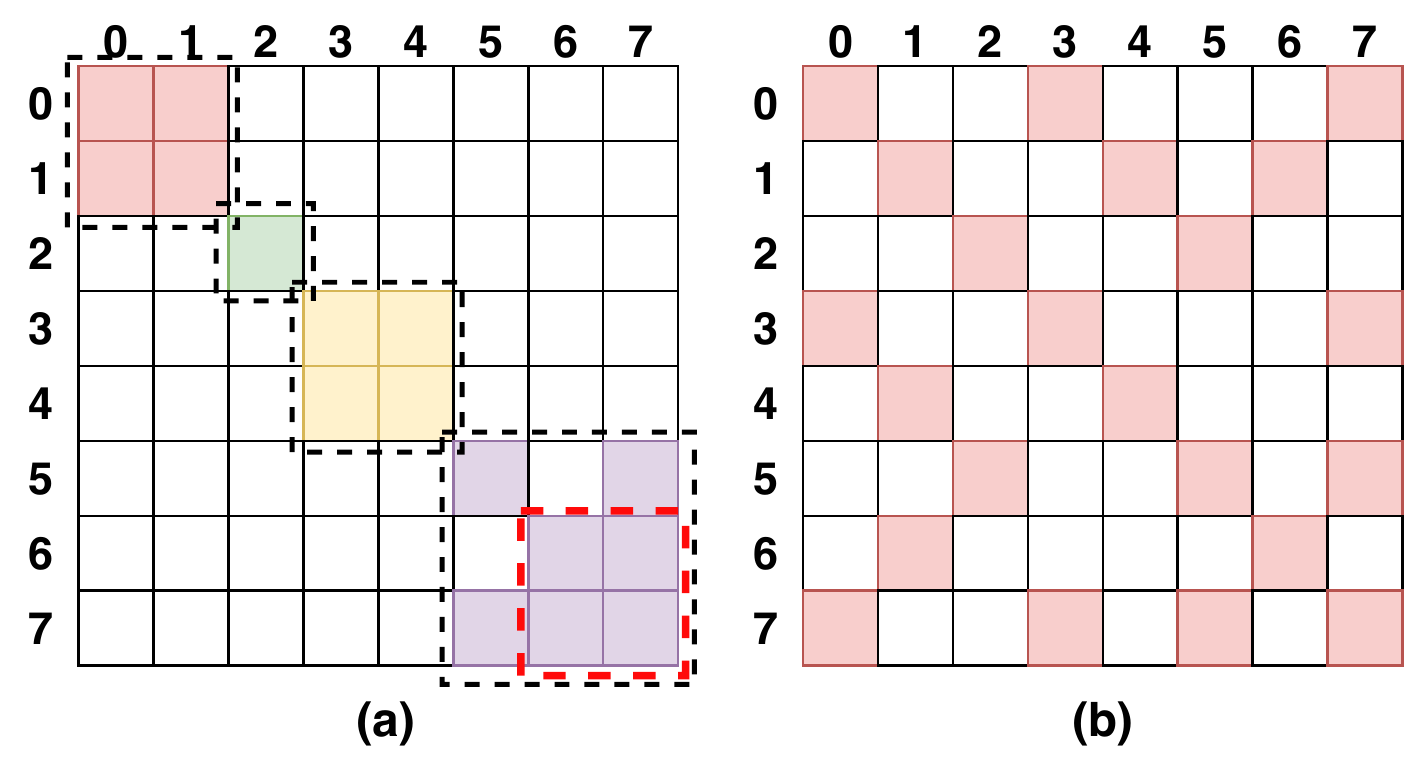}
    \vspace{-2pt}
    \caption{Graph Edge Connection Patterns. Note that each colored square represents the edge between two nodes. Different colors in (a) represent edges from different communities. The red dot-line box indicates the sub-community.}
    \vspace{-10pt}
    \label{fig: node renumbering}
\end{figure}

\textbf{When to apply: } While graph reordering provides potential benefits for performance, we still need to figure out \textit{what kind of graph would benefit from such reordering optimization}. 
Our key insight is that for graphs already in a shape approximating block-diagonal pattern in their adjacency matrix (Figure~\ref{fig: node renumbering}\subfig{a}), reordering could not bring more locality benefits, since nodes within each community are already close to each other in terms of their node-IDs. 
For graphs with a more irregular shape (Figure~\ref{fig: node renumbering}\subfig{b}), where edge connections are distributed among nodes with an irregular pattern, the reordering could bring notable performance improvement (up to 2$\times$ speedup, later discussed in Section~\ref{sect: Additional Studies}). 
To this end, we propose a new metric -- \textit{Averaged Edge Span} (AES), 
to determine whether it is beneficial to conduct a graph reordering.
\begin{equation} \small \label{equ: graph diameter}
    \mathbf{AES} = \frac{1}{\# E} \sum\limits_{(src_{id}, trg_{id}) \in E} |src_{id} - trg_{id}|
\end{equation}
where $E$ is the edge set of the graph; $\#E$ is the number of total edges; $src_{id}$ and $trg_{id}$ are the source and target node IDs of each edge.
Computing AES is lightweight and can be done on-the-fly during the initial graph loading. 
Our profiling of a large corpus of graphs also shows that when $\sqrt{AES} > \lfloor\frac{\sqrt{\#N}}{100}\rfloor$ node numbering is more likely to improve runtime performance. 

\textbf{How to apply: } 
We leverage Rabbit Reordering~\cite{rabbit-order}, which is a fully parallelized and low-cost graph reordering technique. Specifically, it first maximizes the graph modularity by hierarchically merging edges and clustering nodes. And it then generates node order within each cluster through DFS traversal. 
Rabbit Reordering has also been proved to outperform other graph clustering approaches~\cite{METIS, boldi2011layered, raghavan2007near, karantasis2014parallelization, RCM-Algorithm}, including Community-based methods, such as METIS~\cite{METIS}, and BFS-based methods, such as Reverse Cuthill-McKee (RCM)~\cite{RCM-Algorithm}) in terms of better quality (data locality) of the captured graph communities, the ease of parallelization, and performance. 
More importantly, Rabbit Reordering can capture the graph communities hierarchically (\textit{i.e.}, a set of smaller sub-communities are included in a larger community, as exemplified in Figure~\ref{fig: node renumbering}\subfig{a}). 
Such communities at different levels of granularities would be a good match for the GPU cache hierarchy, where smaller sub-communities (occupying one SM) can enjoy the data locality benefit from the L1 cache, while larger communities (occupying multiple SMs) can enjoy the data locality from the larger L2 cache. We quantitatively discuss such a locality benefit in Section~\ref{sect: Optimization Analysis}.

\subsection{Warp-aware Memory Customization} 
Existing works~\cite{pyG, wang2016gunrock} utilize a large number of global memory accesses for reading and writing the embedding and a large number of atomic operations for aggregation (a reduction operation).
However, this approach leads to heavy overhead and fails to exploit the potential benefits from shared memory.
In particular, when aggregating on a target node with $k$ neighbor groups (each has $ngs$ neighbors with $Dim$-Dimensional embeddings) into a $Dim$-dimensional embedding, it involves $O(k\cdot ngs\cdot Dim)$ atomic operations and $O(k\cdot ngs\cdot Dim)$ global memory accesses.

By contrast, we propose a warp-centric shared memory optimization technique.
Our key insight is that by customizing shared memory layout according to the block-level warp organization pattern (Figure~\ref{fig: node renumbering}),
we can significantly reduce the number of atomic operations and global memory access.
First of all, we reserve a shared memory space ($4\times Dim$ bytes for floating-point embeddings) for the target node of each neighbor group (warp), such that the threads from a warp can cache the intermediate results of reduction in shared memory.
Later on, within a thread block, we designate only one warp (called \textit{leader}) for copying the intermediate results of each target node to global memory considering that neighbors of each node can be spread across different warps. The detailed customization procedure is described in Algorithm~\ref{alg: Memory Organization.}. 
Specifically, each warp (maintained in $warpPtr$) has three properties: $nodeSharedAddr$ (a shared memory address for the aggregation result of a neighbor-group), $nodeID$ (the ID of the target node), and $leader$ (a boolean flag indicating whether the current warp is a leader warp for flushing out the result from the shared memory to the global memory). 
The major customization routine (Line $4$ to Line $22$) handles different warps based on their index position relative to thread blocks. 
Note that such a shared memory customization is low-cost and is done only once on-the-fly with the regular graph initialization process before the GPU kernel execution.

In our design, when a target node with $k$ neighbor groups (each has $ngs$ neighbors with $Dim$-dimensional embeddings), it involves $O(Dim)$ atomic operations and $O(Dim)$ global memory accesses.
To this end, we can save the atomic operations and global memory access by $(k\cdot ngs)\times$, thus significantly accelerating the aggregation operations.
Here, we treat $ngs$ as a hyper-parameter to balance memory access efficiency and computation parallelism, and we further discuss its value selection in Section~\ref{sect: analytical modeling}.
\vspace*{-0.2cm}
\begin{algorithm}[t] \small
    \setstretch{1}
    \caption{Warp-aware Memory Customization.}
    \algsetup{linenosize=\tiny}
    \fontsize{9pt}{9pt}\selectfont
    \label{alg: Memory Organization.}
    \hspace{12pt}\comment{Compute \#neighbor-groups (\#warps)}
    \begin{algorithmic}[1]
    \STATE{$\mathit{warpNum} = \mathit{neighborGroups} =$ \textbf{computeGroups}($\mathit{ngs}$);} \\
    \comment{Compute the number of warps per thread block}
    \STATE{$\mathit{warpPerBlock} = $ \textbf{floor}$(\mathit{threadPerBlock}/\mathit{threadPerWarp})$
    } \\
    \comment{Initialize tracking variables}
    \STATE{$\mathit{cnt} = 0$; $\mathit{local\_cnt} = 0$; $\mathit{last} = 0$;} 
    \WHILE {$\mathit{cnt} < \mathit{warpNum}$}{
        \item[\hspace{25pt} \comment{Warp in the front of a thread block}]
        \IF {$\mathit{cnt}$ \% $\mathit{warpPerBlock} == 0$}
        { 
            \STATE{$warpPtr[cnt].\mathit{nodeSharedAddr} = \mathit{local\_cnt} \times \mathit{Dim}$;}\\
            \STATE{$\mathit{last}= \mathit{warpPtr}[cnt].nodeID$;}\\
            \STATE{$\mathit{warpPtr}[cnt].leader = true$;}\\
        }
        \item[\hspace{25pt} \comment{Warp in the middle of a thread block}]
        \ELSE
            { 
            \item[\hspace{35pt} \commentone{Warp with the same target node as}]
            \item[\hspace{42pt} \commenttwo{its predecessor warp}]
            \IF {$\mathit{warpPtr}[cnt].nodeID == last$)}{
                \STATE{$\mathit{warpPtr}[cnt].nodeSharedAddr = local\_cnt$;}
            }
            \item[\hspace{35pt} \commentone{Warp with the different target node as}]
            \item[\hspace{42pt} \commenttwo{its predecessor warp}]
            \ELSE {
                \STATE{$\mathit{local\_cnt}++;$} \\
                \STATE{$\mathit{warpPtr}[cnt].nodeSharedAddr = \mathit{local\_cnt};$} \\
                \STATE{$\mathit{last} = \mathit{warpPtr}[cnt].nodeID;$} \\
                \STATE{$\mathit{warpPtr}[cnt].leader = true;$} \\
            }
            \ENDIF
        }
        \ENDIF
        \item[\hspace{20pt} \comment{Next warp belongs to a new thread block}]
        \IF{$(++\mathit{cnt}) \% \mathit{warpPerBlock} == 0$}{
            \STATE{$\mathit{local\_cnt} = 0;$}\\
        }
        \ENDIF
        }
    \ENDWHILE
    \end{algorithmic}
\end{algorithm}
\section{Design Optimization}
\label{sect: analytical modeling}
\vspace{-5pt}
The parameters in our GPU kernel configurations can be tuned to accommodate various GNN models with graph data sets.
But it is not yet known how to automatically select the parameters which can deliver the optimal performance.
In this section, we introduce the analytical model and the auto parameter selection in the \textbf{\code{Decider}} of \Mname.

\textbf{Analytical Modeling:}
The performance/resource analytical model of \Mname~has two variables, workload per thread ($\mathit{WPT}$), and shared memory usage per block ($\mathit{SMEM}$).
\begin{equation} \small
\label{equ: workload per threads}
\begin{aligned}
    \mathbf{WPT}  = \mathit{ngs}\times\frac{Dim}{\mathit{dw}}, \ \
    \mathbf{SMEM}  =  \frac{\mathit{tpb}}{\mathit{tpw}} \times Dim \times \mathit{FloatS}
\end{aligned}
\end{equation}
where $\mathit{ngs}$ and $\mathit{dw}$ is the neighbor-group and dimension-worker size (Section~\ref{sect: Fine-grained Dimension Partitioning}), respectively; 
$\mathit{Dim}$ is the node embedding dimension; 
$\mathit{IntS}$ and $\mathit{FloatS}$ are both 4-byte on GPUs;
$\mathit{tpb}$ is the thread-per-block and $\mathit{tpw}$ is the thread-per-warp; 
$\mathit{tpw}$ is 32 for GPUs, while $\mathit{tpb}$ is selected by users.

\textbf{Parameter Auto Selection:}
To determine the value of the $\mathit{ngs}$ and $\mathit{dw}$, we follow two steps.
First, we determine the value of $\mathit{dw}$ based on $\mathit{tpw}$ (hardware constraint) and $\mathit{Dim}$ (input property), as shown in Equation~\ref{eq: Dimension-worker selection}. 
Note that we develop this equation by profiling different datasets and GNN models.
\begin{equation} \label{eq: Dimension-worker selection}
   \mathit{dw} = 
    \begin{cases}
     \mathit{tpw}          &   \mathit{Dim} \geq \mathit{tpw} \\ 
    \frac{\mathit{tpw}}{2} &   \mathit{Dim} < \mathit{tpw}
    \end{cases}
\end{equation} 
Second, we determine the value of $\mathit{ngs}$ based on the selected $\mathit{dw}$ and the user-specified $\mathit{tpb}$. 
The constraints include making $\mathit{WPT} \approx 1024$ and $\mathit{SMEM} \leq \mathit{SMEMperBlock}$. Note that $SMEMperBlock$ is $48$KB to $96$KB on modern GPUs~\cite{tesla-v100,quardo}. 
Across different GPUs, even though the number of CUDA cores and global memory bandwidth would be different, the single-thread workload capacity (measured by $\mathit{WPT}$) remains similar.  
$\mathit{tpb}$ is usually chosen as a power of 2 but less than or equal 1024. 
Our insight based on micro-benchmarking and previous literature~\cite{merge-spmm} shows that smaller blocks (1 to 4 warps, \textit{i.e.}, $32 \leq \mathit{tpb} \leq 128$) can improve SM warp scheduling flexibility and avoid tail effects, thus leading to higher GPU occupancy and throughput.
We further demonstrate the effectiveness of our analytical model in Section~\ref{sect: Additional Studies}.

\vspace{-10pt}
\section{Evaluation}
\vspace{-5pt}
In this section, we comprehensively evaluate \Mname~in terms of the performance and adaptability on various GNN models, graph datasets, and GPUs. 
\vspace{-5pt}
\subsection{Experiment Setup}
\hspace{5pt} \textbf{Benchmarks: }
We choose the two most representative GNN models widely used by previous work~\cite{wang2019dgl,pyG,ma2019neugraph} on node classification tasks to cover different types of aggregation.
\underline{1) Graph Convolutional Network (GCN)}~\cite{GCNConv} is one of the most popular GNN model architectures. 
It is also the key backbone network for many other GNNs, such as GraphSAGE~\cite{SageConv}, and differentiable pooling (Diffpool)~\cite{diffpool}. Therefore, improving the performance of GCN will also benefit a broad range of GNNs. For GCN evaluation, we use the setting: \textit{2 layers with 16 hidden dimensions}, which is also the setting from the original paper~\cite{GCNConv}.
\underline{2) Graph Isomorphism Network (GIN)}~\cite{GINConv}.
GIN differs from GCN in its aggregation function, which weighs the node embedding values from the node itself. In addition, GIN is also the reference architecture for many other advanced GNNs with more edge properties, such as Graph Attention Network (GAT)~\cite{GATConv}. For GIN evaluation, we use the setting: \textit{5 layers with 64 hidden dimensions}, which is the setting used in the original paper~\cite{GINConv}.

\textbf{Baselines: } 
we choose several baseline implementations for comparison. 
\underline{1) {Deep Graph Library (DGL})}~\cite{wang2019dgl} is the state-of-the-art GNN framework on GPUs, which is built upon the famous tensor-oriented platform -- Pytorch~\cite{pytorch}. DGL significantly outperforms the other existing GNN frameworks~\cite{pyG} over various datasets on many mainstream GNN architectures. Therefore, we make an in-depth comparison with DGL in our evaluation;
\underline{2) {Pytorch-Geometric (PyG})}~\cite{pyG} is another
GNN framework in which users can define their edge convolutions when building customized GNN aggregation layers; 
\underline{3) {NeuGraph}}~\cite{ma2019neugraph} is a dataflow-centered GNN system on GPUs built on Tensorflow~\cite{tensorflow2015}; 
\underline{4) {Gunrock}}~\cite{wang2016gunrock} is the GPU-based graph processing framework with state-of-the-art performance on traditional graph algorithms (\textit{e.g.}, PageRank). 

\textbf{Datasets: }
We cover all three types of datasets, which have been used in previous GNN-related work~\cite{wang2019dgl, pyG, ma2019neugraph}.
\underline{Type I} graphs are the typical datasets used by previous GNN algorithm papers~\cite{GCNConv, GINConv, SageConv}. 
They are usually small in the number of nodes and edges, but rich in node embedding information with high dimensionality. 
\underline{Type II} graphs~\cite{KKMMN2016} are the popular benchmark datasets for graph kernels and are selected as the built-in datasets for PyG~\cite{pyG}. Each dataset consists of a set of small graphs, which only have intra-graph edge connections without inter-graph edge connections. 
\underline{Type III} graphs~\cite{snapnets, GCNConv} are large in terms of the number of nodes and edges. These graphs demonstrate high irregularity in structure, which is challenging for most of the existing GNN frameworks. Details of these datasets are listed in Table~\ref{table: Evaluation Dataset}. 
\begin{table}[t] \footnotesize	
\caption{Datasets for Evaluation.}
\vspace{-6pt}
\centering
\scalebox{0.94}{
 \begin{tabular}{|| c | l r r r r ||}
 \hline
\textbf{Type} & \textbf{Dataset} & \textbf{\#Vertex} & \textbf{\#Edge} & \textbf{Dim.} & \textbf{{\#Class}}\\
\hline
\multirow{4}{*}{\textbf{I}} & Citeseer    & 3,327	    & 9,464	    & 3,703 & 6      \\
& Cora	    & 2,708     & 10,858	& 1,433 & 7      \\
& Pubmed	    & 19,717	& 88,676	& 500  & 3      \\
& PPI	        & 56,944	& 818,716	& 50   & 121    \\
\hline
\hline

\multirow{6}{*}{\textbf{II}} 
& PROTEINS\_full	&   43,471	       & 162,088	&   29	    & 2 \\
& OVCAR-8H	    &   1,890,931	   & 3,946,402	&   66	    & 2 \\
& Yeast	        &   1,714,644	   & 3,636,546	&   74	    & 2 \\
& DD	            &   334,925	       & 1,686,092	&   89	    & 2 \\
& TWITTER-Partial	&   580,768	       & 1,435,116	&   1,323    & 2 \\
& SW-620H	        &   1,889,971	   & 3,944,206	&   66	    & 2 \\
\hline
\hline

\multirow{5}{*}{\textbf{III}} 
& amazon0505	    & 410,236	& 4,878,875	    & 96  & 22 \\
& artist	        & 50,515	& 1,638,396	    & 100 & 12 \\
& com-amazon	    & 334,863	& 1,851,744	    & 96  & 22 \\
& soc-BlogCatalog	& 88,784	& 2,093,195	    & 128 & 39 \\
& amazon0601	    & 403,394	& 3,387,388	    & 96 & 22 \\
\hline
\end{tabular}}
\label{table: Evaluation Dataset}
\end{table}

\textbf{Platforms \& Metrics: } 
\label{sect: Platforms and Metrics }
We implement \Mname's backend with C++ and CUDA C and its front-end with Python. Our major evaluation platform is a server with an 8-core 16-thread Intel Xeon Silver 4110 CPU~\cite{xeon} 
and a Quadro P6000~\cite{quardo} GPU.
Besides, we use Tesla V100~\cite{tesla-v100} GPU on the DGX-1 system~\cite{dgx} to demonstrate the generality of \Mname. 
Runtime parameters of different input settings are optimized by \Mname~\code{\textbf{Decider}}. 
To measure the performance speedup, we calculate the averaged latency of 200 end-to-end inference (forward propagation) or training (forward+backward propagation).

\subsection{Compared with DGL} \label{sect: compared with DGL}
\begin{figure} [t] \small
    \centering
    \includegraphics[width=\columnwidth]{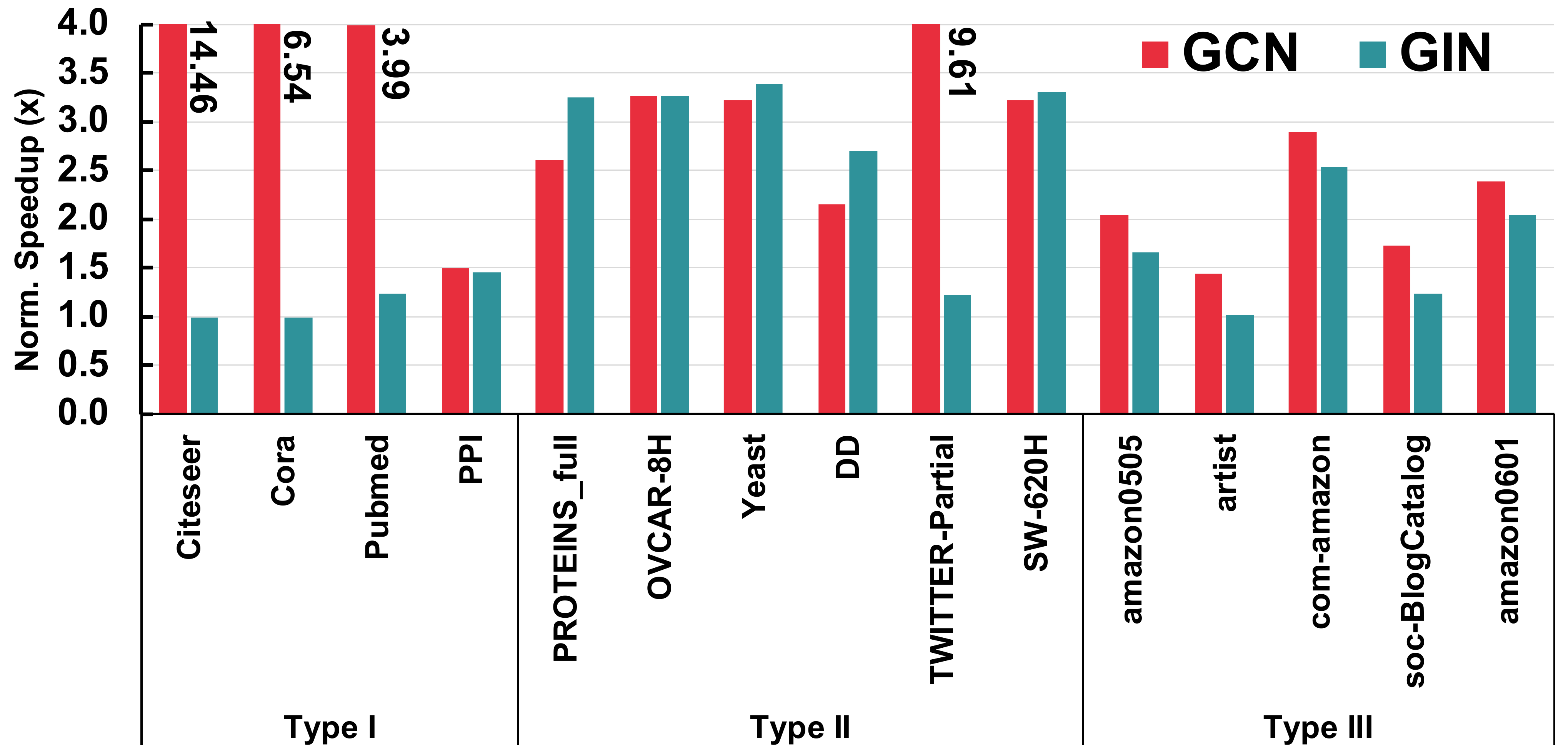}
    \vspace{-15pt}
    \caption{Inference speedup ($\times$) over DGL on GCN and GIN.}
    \vspace{-5pt}
    \label{fig: Speedup vs DGL.}
\end{figure}

In this section, we first conduct a detailed experimental analysis and comparison with DGL on GNN inference, then extend our comparison for GNN training. 
As shown in Figure~\ref{fig: Speedup vs DGL.}, \Mname~achieves $4.03\times$ and $2.02\times$ speedup on average compared to DGL~\cite{wang2019dgl} over three types of datasets for GCN and GIN on inference, respectively. 
We next provide detailed analysis and give insights for each type of datasets.

\textbf{Type I Graphs:}  
The performance improvement against DGL is significantly higher for GCN (on average $6.45\times$) than GIN (on average $1.17\times$). 
The major reason is their different GNN computation patterns. 
For GCN, node dimension reduction (DGEMM) is always placed before aggregation. This largely reduce data movement and thread synchronization overheads during the aggregation phase, 
which could gain more benefits from \Mname's 2D workload management and specialized memory optimization for data locality improvements. GIN, on the other side, has aggregation phase that must be finished before the node dimension reduction. Thus, it cannot avoid high-volume memory access and data movements during the aggregation phase. 
Therefore, it gets lower benefits from the data locality and the shared memory on GPUs for fast and low-overhead memory access. 
However, our fine-grained dimension partitioning can still handle these high-dimensional cases effectively.


\textbf{Type II Graphs:} Performance shows less difference between GCN ($4.02\times$) and GIN ($2.86\times$) on the same datasets except for \textit{TWITTER-Partial}, which has the highest node embedding dimension (1323) in Type II graphs. It is worth noticing that the speedup for GIN is consistently better compared with Type I. There are two major reasons: 1) node feature dimension is much lower (average 66.5, excluding \textit{TWITTER-Partial}) versus Type I (average 1421), which can gain more performance benefits from data spatial and temporal locality of our specialized memory optimizations; 2) Type II graphs intrinsically have good locality in their graph structure. The reason is that Type II datasets consist of small graphs with very dense intra-graph connections but no inter-graph edges, plus nodes within each small graph are assigned with consecutive IDs. Therefore, the performance gains of such graph-structure locality can be scaled up when combining with \Mname's efficient workload and memory optimizations.

\vspace{-1pt}
\textbf{Type III Graphs:} The speedup is also evident (average $2.10\times$ for GCN and average $1.70\times$ for GIN) on graphs with a large number of nodes and edges, such as \textit{amazon0505}. The reason is the high overhead inter-thread synchronization and global memory access can be well reduced through our 2D workload management and specialized memory optimization. 
Besides, our community-aware node renumbering further facilitates an efficient workload sharing among adjacent threads (working on a group of nodes) through improving the data spatial/temporal locality. 
On the dataset \textit{artist}, which has the smallest number of nodes and edges within Type III, we notice a lower performance speedup for GIN. And we find that the \textit{artist} dataset has the highest standard deviation of graph community sizes within Type III graphs, which makes it challenging to 1) use the group community information to capture the node temporal and spatial locality in the GNN aggregation phase, and 2) capitalize on the performance benefits of using such a community structure for guiding system-level optimizations (\textit{e.g.},
warp-aligned thread mapping and shared memory customization) on GPUs, which have a fixed number of computation and memory units within each block/SM. 

\textbf{Kernel Metrics:} 
For detailed kernel metrics analysis, we utilize NVProf~\cite{nvprof} to measure two performance-critical (computation and memory) CUDA kernel metrics: \textit{Stream Processor (SM) efficiency} and \textit{Cache (L1 + L2 + Texture) Hit Rate}.
\Mname~achieves on average $24.47\%$ and $12.02\%$ higher SM efficiency compared with DGL for GCN and GIN, respectively, which indicates that our 2D workload management can strike a good balance between the single-thread efficiency and the multi-thread parallelism that are crucial to the overall performance improvement. 
\Mname~achieves on average $75.55\%$ and $126.20\%$ better cache hit rate compared with DGL for GCN and GIN, correspondingly, which demonstrates the benefit of specialized memory optimizations.
\begin{figure} [t] \small
    \centering
    \includegraphics[width=\columnwidth, height=4cm]{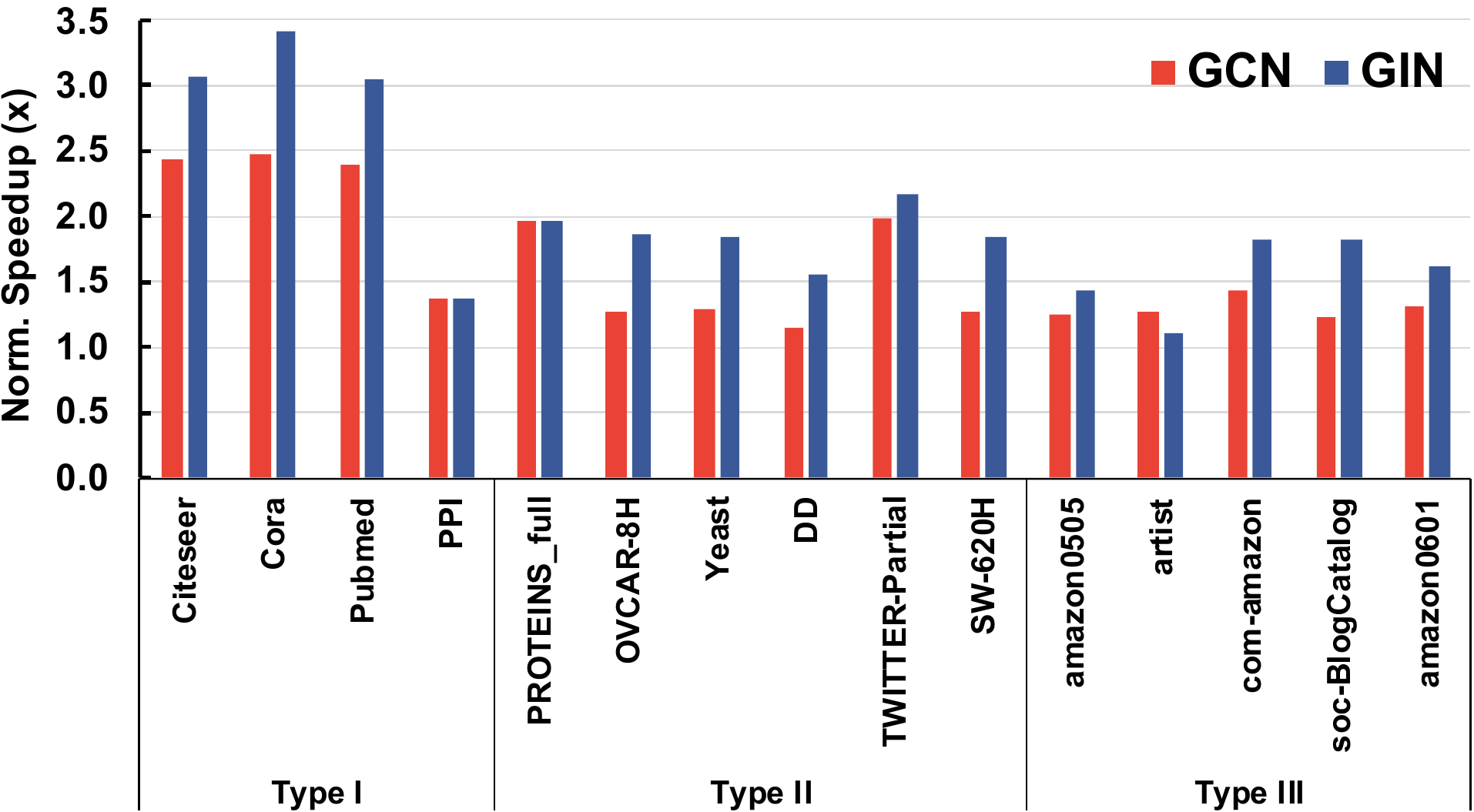}
    \vspace{-15pt}
    \caption{Training speedup ($\times$) over DGL on GCN and GIN.}
    \vspace{-5pt}
    \label{fig: GCN Training Performance Comparison with DGL.}
\end{figure}

\vspace{-10pt}
\textbf{Training Support:}
We also evaluate the training performance of \Mname~on all three types of datasets compared with the DGL on both GCN and GIN. 
Compared with inference, training is more challenging, since it involves more intensive computation with the forward value propagation and the backward gradient propagation, both of which heavily rely on the underlying graph aggregation kernel for computation. 
As shown in Figure~\ref{fig: GCN Training Performance Comparison with DGL.}, \Mname~consistently outperforms the DGL framework with average $1.61\times$ and average $2.00\times$ speedup on GCN and GIN, respectively, which shows the strength of our input-driven optimizations.
The key difference between training and inference of GNNs is two-fold: First, backpropagation is needed in training. This step benefits from our improvements, as the backpropagation step is similar to the forward computation during the inference, and all the proposed methods are still beneficial; Second, training incurs extra memory and data movement overheads for storing/accessing the activations of the forward pass until gradients can be propagated back. 

\subsection{Compared with other Frameworks}
We compare with DGL on all input settings, since DGL is the overall best-performance GNN framework.
In this section, we further compare \Mname~with three other representative GNN computing frameworks on their best settings.
\begin{figure}[t] \small
    \centering
    \includegraphics[width=0.48\textwidth, height=2.3cm]{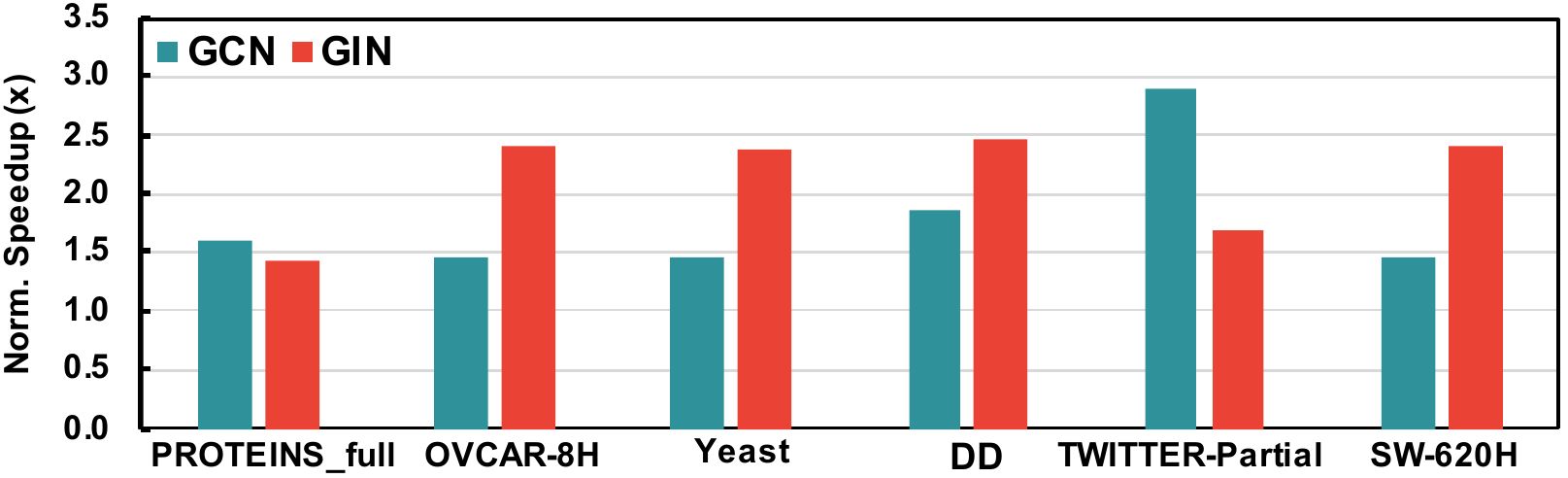}
    \vspace{-15pt}
    \caption{Training speedup ($\times$) over PyG on GCN and GIN.}
    \vspace{-5pt}
    \label{fig: Comparison with PyG.}
\end{figure}

\textbf{Compared with PyG: } 
As shown in Figure~\ref{fig: Comparison with PyG.}, \Mname~can outperform PyG with $1.78\times$ and $2.13\times$ speedup on average for GCN and GIN, respectively. For GCN, \Mname~achieves significant speedup on datasets with high-dimensional node embedding, such as \textit{TWITTER-Partial}, through 1) node dimension reduction before aggregation and 2) workload sharing among neighbor partitions and dimension partitions. For GIN, \Mname~reaches $2.45\times$ speedup on datasets with a higher average degree, such as \textit{DD}, since \Mname~can effectively distribute the workload of each node along their embedding dimension to working threads while balancing the single-thread efficiency and inter-thread parallelism. PyG, however, achieves inferior performance because 1) it has poor thread management in balancing workload and controlling synchronization overhead; 2) it heavily relies on the scatter-and-gather kernel, which lacks flexibility. 
\begin{table}[t] \small	
\centering
\caption{Latency (ms) comparison with NeuGraph (NeuG).}
\vspace{-5pt}
\scalebox{0.96}{
\begin{tabular}{|l|r|r|r|r|}
\hline
\textbf{Dataset}   & \textbf{NeuG (ms)} & \textbf{Ours~(ms)}  & \textbf{Speedup}\\ 
\hline
\hline
reddit-full          & 2460   & 599.69  & 4.10$\times$              \\ \hline
enwiki               & 1770   & 443.00  & 3.99$\times$             \\ \hline
amazon               & 1180   & 474.57  & 2.48$\times$              \\ \hline
\end{tabular}}
\label{tbl: Comparison with Neugraph.}
\vspace{-10pt}
\end{table}
\begin{figure} [t] \small
    \centering
    \includegraphics[width=\columnwidth]{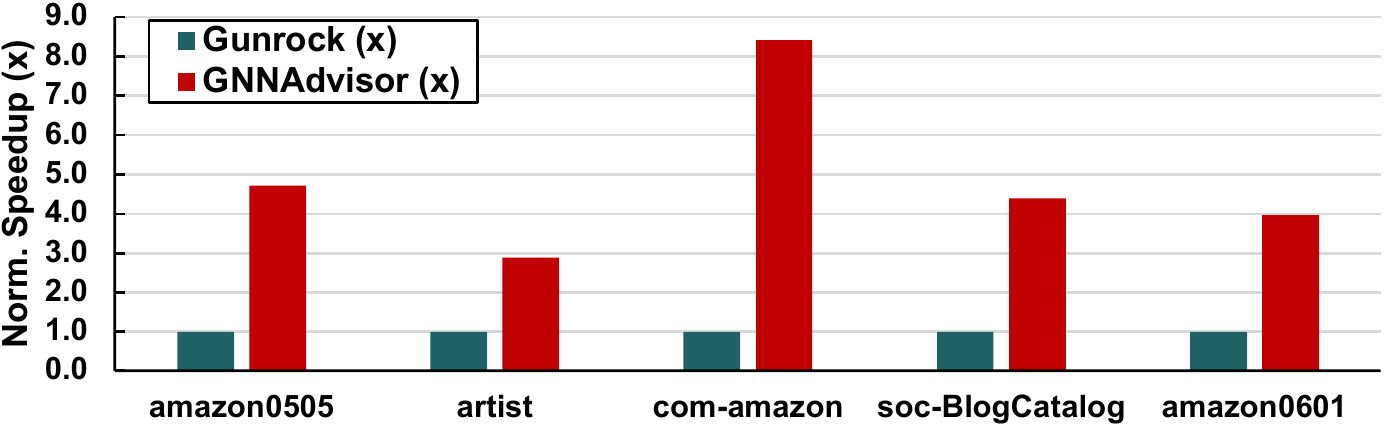}
    \vspace{-10pt}
    \caption{Speedup ($\times$) comparison with Gunrock.}
    \vspace{-15pt}
    \label{fig: Comparison with Gunrock.}
\end{figure}

\begin{figure*}[t] \small
    \centering
    \subfloat[]{\includegraphics[width=0.25\textwidth, height=1.9cm, trim=0 0.6cm 0 0 ]{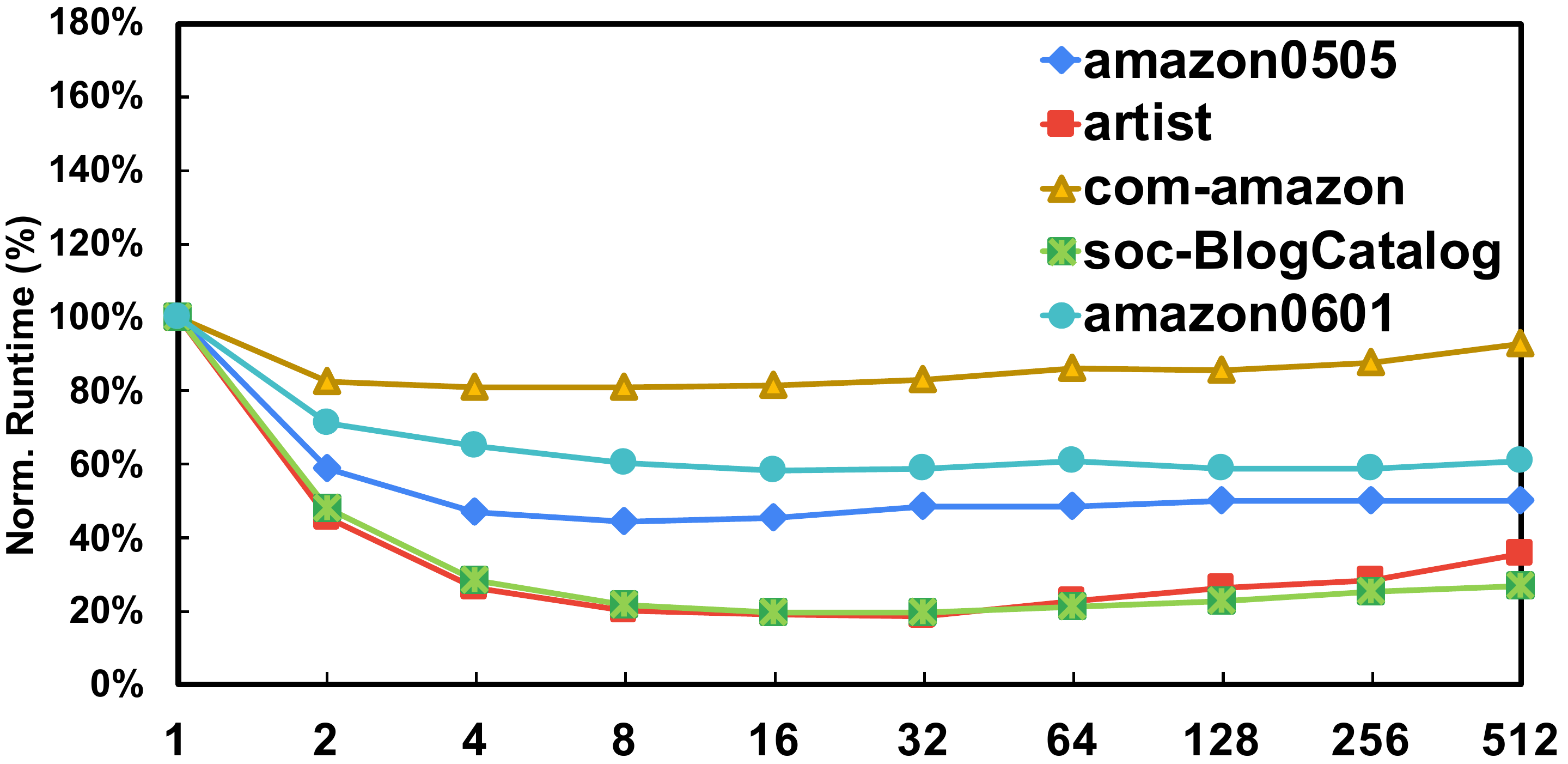}}
    \subfloat[]{\includegraphics[width=0.25\textwidth, trim=0 0.4cm 0 0cm, height=1.9cm]{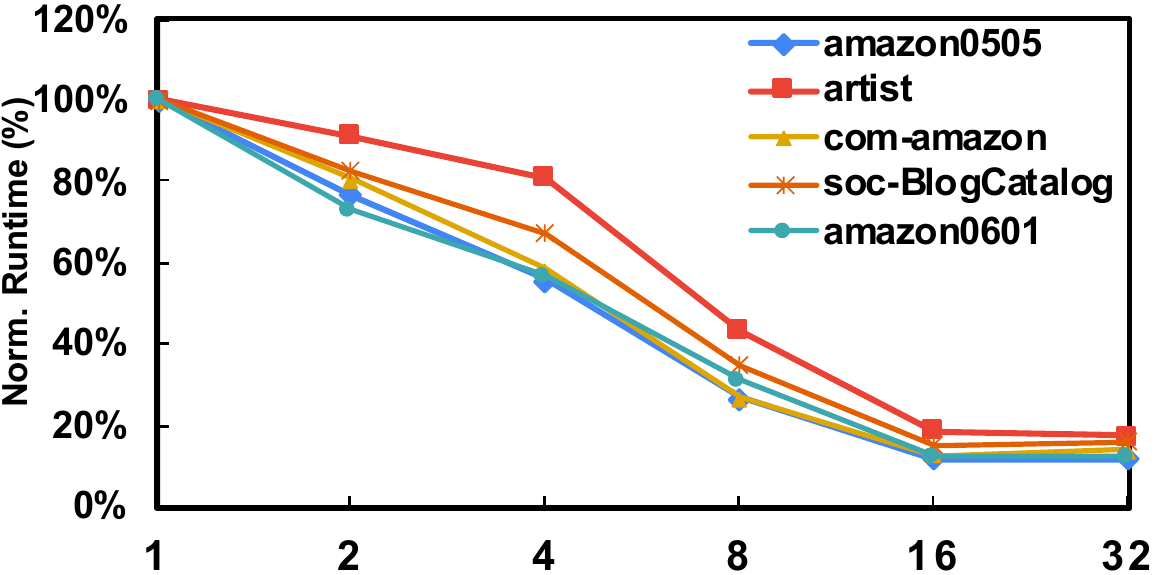}}
    \subfloat[]{\includegraphics[width=0.25\textwidth, height=1.9cm, trim=0cm 0.6cm 0 0]{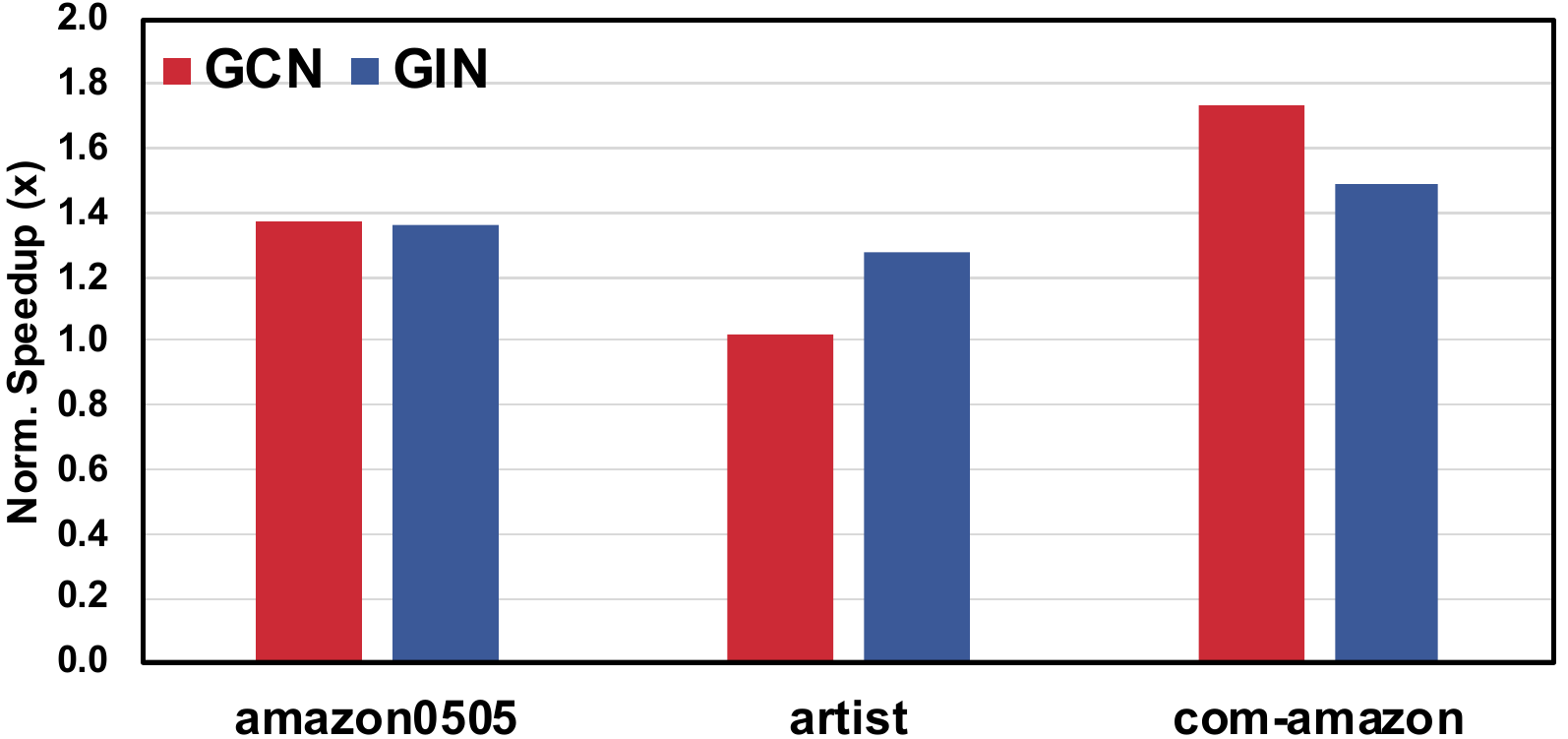}}
    \subfloat[]{\includegraphics[width=0.25\textwidth, height=1.9cm, trim=0 0.4cm 0 0]{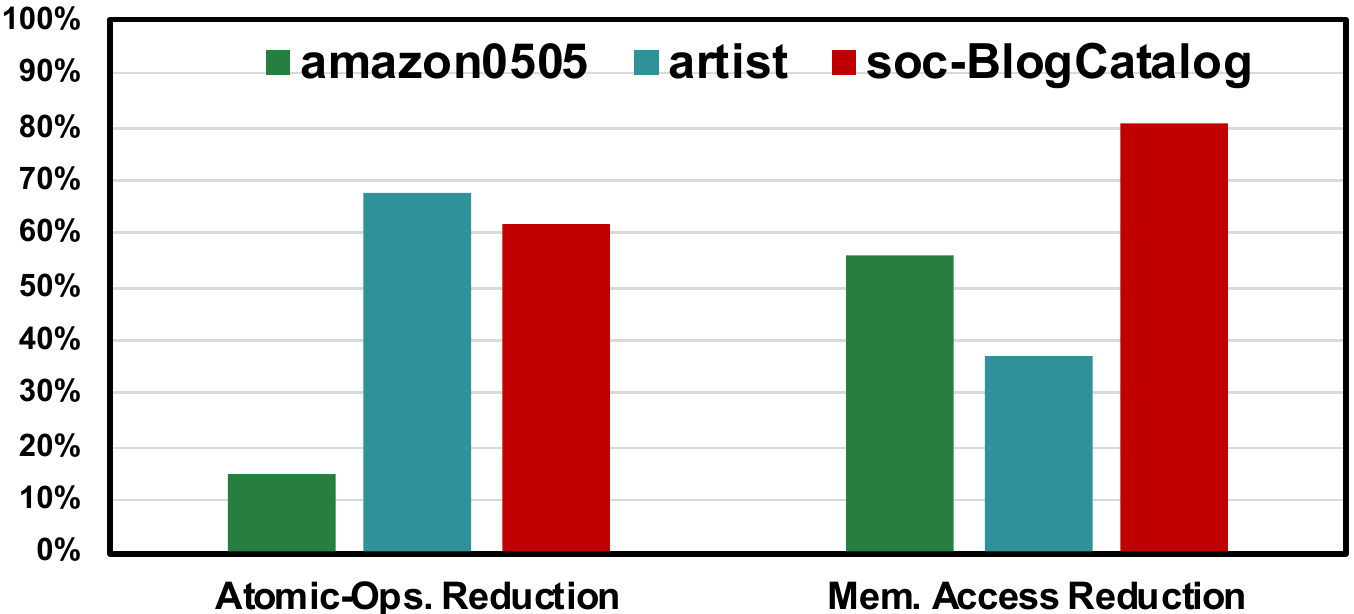}}
    
    \vspace{-10pt}
    \caption{Optimization Analysis. (a) Normalized latency as the neighbor group size ($\mathit{ngs}$) grows (latency at $\mathit{ngs}=1$ is set as $100\%$); (b) Normalized latency as the number of dimension workers grows (latency at $\mathit{dw}=1$ is set as $100\%$); (c) Normalized speedup when using node renumbering compared to without renumbering; (d) Normalized GPU kernel metrics when using block-level optimizations compared to without block-level optimizations.
    }
    \label{fig: Optimization Analysis}
    \vspace{-5pt}
\end{figure*}

\textbf{Compared with NeuGraph: }
For a fair end-to-end training comparison with NeuGraph that has not open-sourced its implementation and datasets,
we 1) use the GPU (Quadro P6000~\cite{quardo}) that is comparable with the GPU of NeuGraph (Tesla P100~\cite{tesla-p100}) in performance-critical factors, such as GPU architecture (both have the Pascal architecture) and the number of CUDA cores; 
2) use the same set of inputs as NeuGraph on the same GNN architecture~\cite{ma2019neugraph};
3) use the datasets that are presented in their paper and are also publicly available.
As shown in Table~\ref{tbl: Comparison with Neugraph.}, \Mname~outperforms NeuGraph with a significant amount of margin ($1.3\times$ to $7.2\times$ speedup) in terms of computation and memory performance. 
NeuGraph relies on general GPU kernel optimizations and largely ignores the input information.
Moreover, the optimizations in NeuGraph are built-in and fixed inside the framework without performance tuning flexibility.
In contrast, \Mname~leverages GNN-featured GPU optimizations and demonstrates the key contribution of input insights for system optimizations. 

\textbf{Compared with Gunrock: }
We make a performance comparison between \Mname~and Gunrock~\cite{wang2016gunrock} on a single neighbor aggregation kernel of GNNs (\textit{i.e.}, the Sparse-Matrix Dense-Matrix Multiplication (SpMM)) over the Type III graphs. 
%
As shown in Figure~\ref{fig: Comparison with Gunrock.}, \Mname~outperforms Gunrock with $2.89\times$ to $8.41\times$ speedup. 
There are two major reasons behind such a evident performance improvement on the sparse GNN computation:
1) Gunrock focuses on graph-algorithm operators (\textit{e.g.}, frontier processing) but lacks efficient support for handling high-dimensional node embedding;
2) Gunrock leverages generic optimizations without considering the input differences, thus, losing the adaptability for handling different GNN inputs efficiently.

\vspace{-5pt}
\subsection{Optimization Analysis} 
\label{sect: Optimization Analysis}
In this section, we explore and analyze the optimizations used in Sections~\ref{sect: 2D Workload Management} and \ref{sect: Specialized Memory Optimization} in detail. 


\textbf{{Neighbor partitioning:}} 
From Figure~\ref{fig: Optimization Analysis}\subfig{a}, we can see that with the increase of the neighbor-group size, the running time of \Mname~will first decrease. The increase of the neighbor-group size saturates the computation capability of each thread meanwhile improving the data locality and reducing the number of atomic operations (\textit{i.e.}, inter-thread synchronization overhead). However, when the neighbor-group size becomes larger than a certain threshold (\textit{e.g.}, 32 for the \textit{artist} dataset), each thread reaches its computation capacity upper bound, and further increasing the neighbor-group size offers no more performance benefit instead increases the overall latency.

\textbf{Dimension partitioning:} 
As shown in Figure~\ref{fig: Optimization Analysis}\subfig{b}, the dimension worker impact 
is more evident in performance compared with the neighbor-group size at the range from 1 to 16. 
When the number of dimension worker increases from 16 to 32, the runtime performance shows very minor difference due to the already balanced single-worker efficiency and multi-worker parallelism. Therefore, further increase the number of dimension workers brings no more benefits.
\begin{figure*}[t] \small
    \vspace{-8pt}
    \subfloat[]{\includegraphics[width=0.33\textwidth, height=2.4cm, trim=0 0.3cm 0 0]{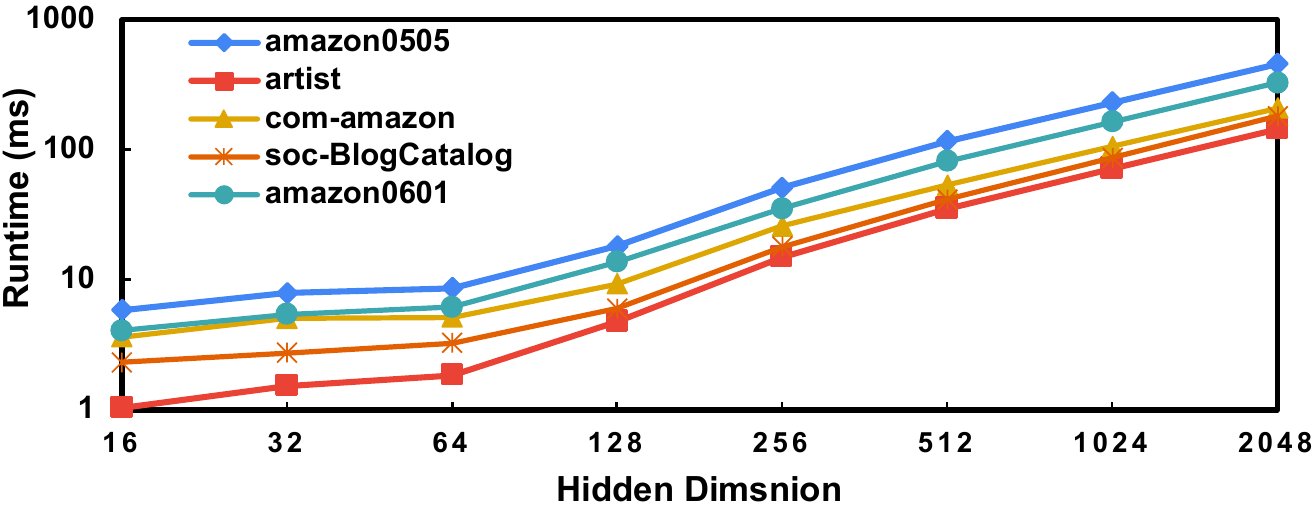}}
    \subfloat[]{\includegraphics[width=0.33\textwidth, height=2.4cm, trim=0 -0.1cm 0 -0.1cm]{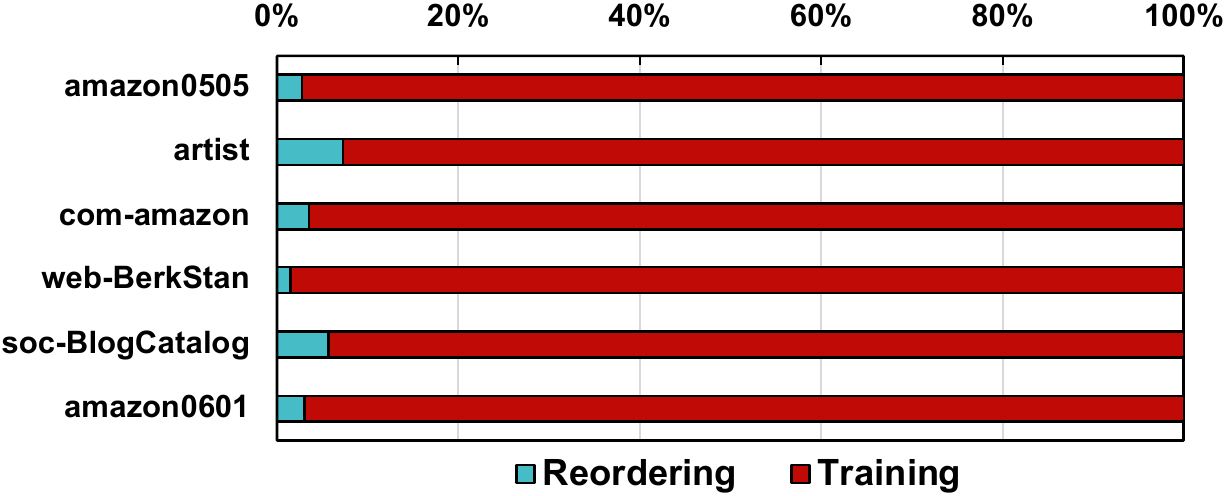}}
    \subfloat[]{\includegraphics[width=0.33\textwidth, height=2.4cm]{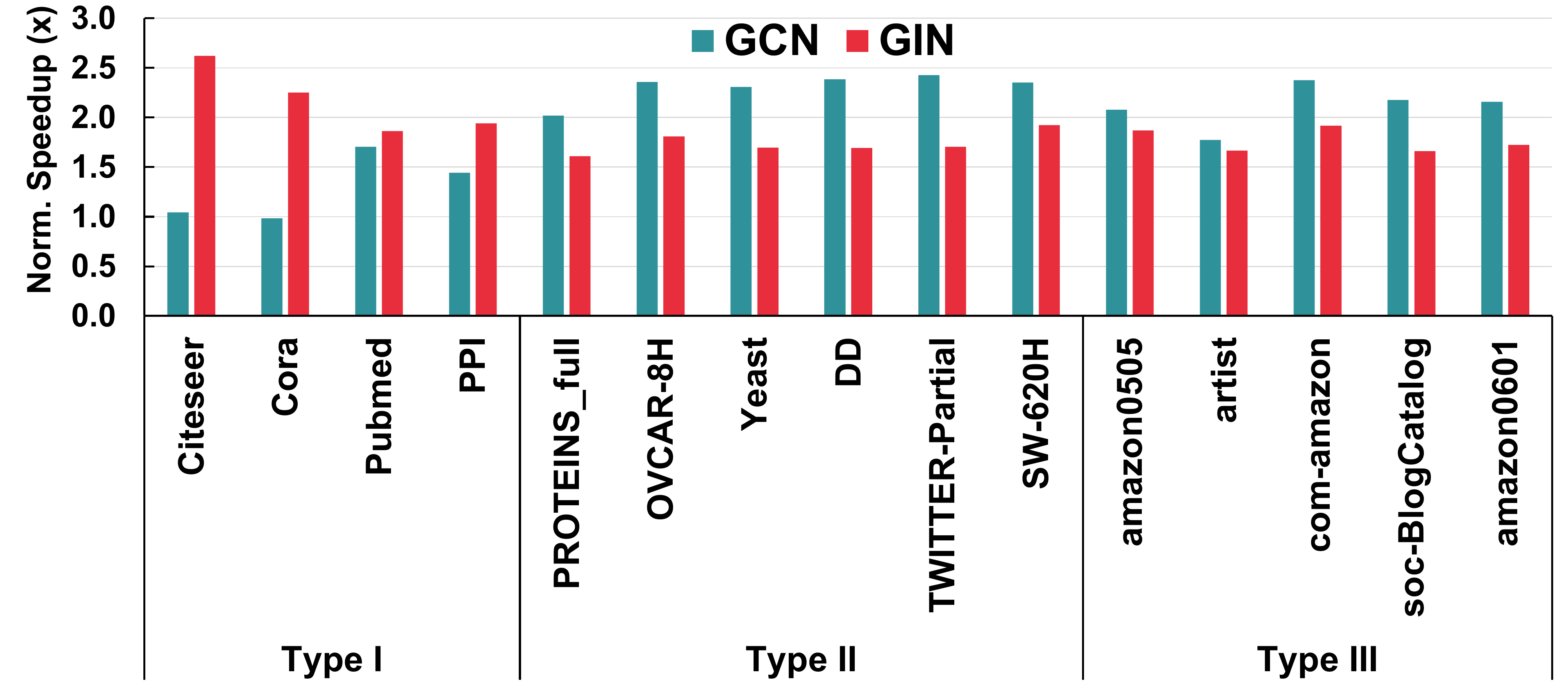}}
\vspace{-10pt}
\caption{Additional Studies. (a) Latency ($ms$) analysis as the hidden dimension grows on GCN; (b) Overhead (\%) analysis for node renumbering; (c) Speedup ($\times$) on Tesla V100 over Quadro P6000 (set as 1$\times$).}
\vspace{-10pt}
\label{fig: Additional Studies.}
\end{figure*}

\textbf{Node renumbering:}
We demonstrate the benefit of node renumbering by profiling Type III datasets for GCN and GIN. As shown in Figure~\ref{fig: Optimization Analysis}\subfig{c}, renumbering nodes within a graph can bring up to $1.74\times$ and $1.49\times$ speedup for GCN and GIN, respectively. The major reason is that our community-aware node renumbering can increase the data spatial and temporal locality during GNN aggregation.

To quantify such locality benefits, we extract the detailed GPU kernel metric -- memory access in terms of read and write bytes from DRAM for illustration. 
Our CUDA kernel metric profiling results show that node renumbering can effectively reduce the memory access overhead (on average 40.62\% for GCN and 42.33\% for GIN) during the runtime since more loaded node embeddings are likely to be shared among the nodes with consecutive IDs. 
We also notice one input case that benefits less from our optimization -- \textit{artist}, since 1) the community size inside \textit{artist} displays a large variation (high standard deviation), making it challenging to capture the neighboring adjacency and locality; 2) such a variation hurdles system-level (computation and memory) optimizations to effectively capitalize on the locality benefits of renumbering. 

\textbf{Block-level optimization: }
We show the optimization benefits of our block-level optimization (including warp-aligned thread mapping, and warp-aware shared memory customization). 
We analyze two kernel metrics (\textit{atomic operations reduction} and \textit{DRAM access reduction}) on three large graphs for illustration. 
As shown in Figure~\ref{fig: Optimization Analysis}\subfig{d}, \Mname~can effectively reduce the atomic operations and DRAM memory access by an average 47.85\% and 57.93\%. 
This result demonstrates 
1) warp-aligned thread mapping based on neighbor partitioning can effectively reduce a large portion of atomic operations; 
2) warp-aware shared memory customization can avoid a significant amount of global memory access.

\vspace{-6pt}
\subsection{Additional Studies} 
\label{sect: Additional Studies}
\hspace{5pt} \textbf{Hidden dimensions of GNN: } 
In this experiment, we analyze the impact of the GNN architecture in terms of the size of the hidden dimension for GCN and GIN. 
As shown in Figure~\ref{fig: Additional Studies.}\subfig{a}, we observe that with the increase of hidden dimension of GCN, the running time of \Mname~is also increased due to more computation (\textit{e.g.}, additions) and memory operations (\textit{e.g.}, data movements) during the aggregation phase and a larger size of the node embedding matrix during the node update phase. 
Meanwhile, we also notice that GIN shows a larger latency increase versus GCN, mainly because of the number of layers (2-layer GCN \textit{vs.} 5-layer GIN) that make such a difference more pronounced.

\textbf{Overhead analysis: }
Community-aware node renumbering is the major source of overhead for leveraging GNN input information, and other parts are negligible. 
Here as a case study, we evaluate its overhead on the training phase of GCN on Type III graphs, given the optimization decision from our \Mname~\textbf{\code{Decider}} (as discussed in Section~\ref{sect: Specialized Memory Optimization}). 
Here we use training for illustration; inference in a real GNN application setting would also use the same graph structure many times~\cite{SageConv, GCNConv, GCNConv} with different node embeddings inputs.
As shown in Figure~\ref{fig: Additional Studies.}\subfig{b}, node-renumbering overhead is consistently small (average 4.00\%) compared with overall training time. 
We thus conclude that such one-time overhead can be amortized over GNN running time, which demonstrates its applicability in real-world GNN applications.

\textbf{Performance on Tesla V100: }
To demonstrate the potential of \Mname~in the modern data-center environment, we showcase the performance of \Mname~on an enterprise-level GPU -- Tesla V100~\cite{tesla-v100}. 
As shown in Figure~\ref{fig: Additional Studies.}\subfig{c}, \Mname~can scale well towards such a high-end device, which can achieve $1.97\times$ and $1.86\times$ speedup compared with P6000 for GCN and GIN due to more computation resources (\textit{e.g.}, 2.6$\times$ SMs, and $1.33\times$ CUDA cores, and $1.13\times$ throughput performance) and higher memory bandwidth (\textit{e.g.}, $2.08\times$ peak memory bandwidth). 
This comparison shows that \Mname~well adapts towards more advanced GPU hardware for seeking better performance. We also foresee that our current work of \Mname~can be extended to the multi-GPU or distributed data center, benefiting overall performance by improving single GPU efficiency. 
\begin{figure} [t] \small
    \centering
    \includegraphics[width=\columnwidth]{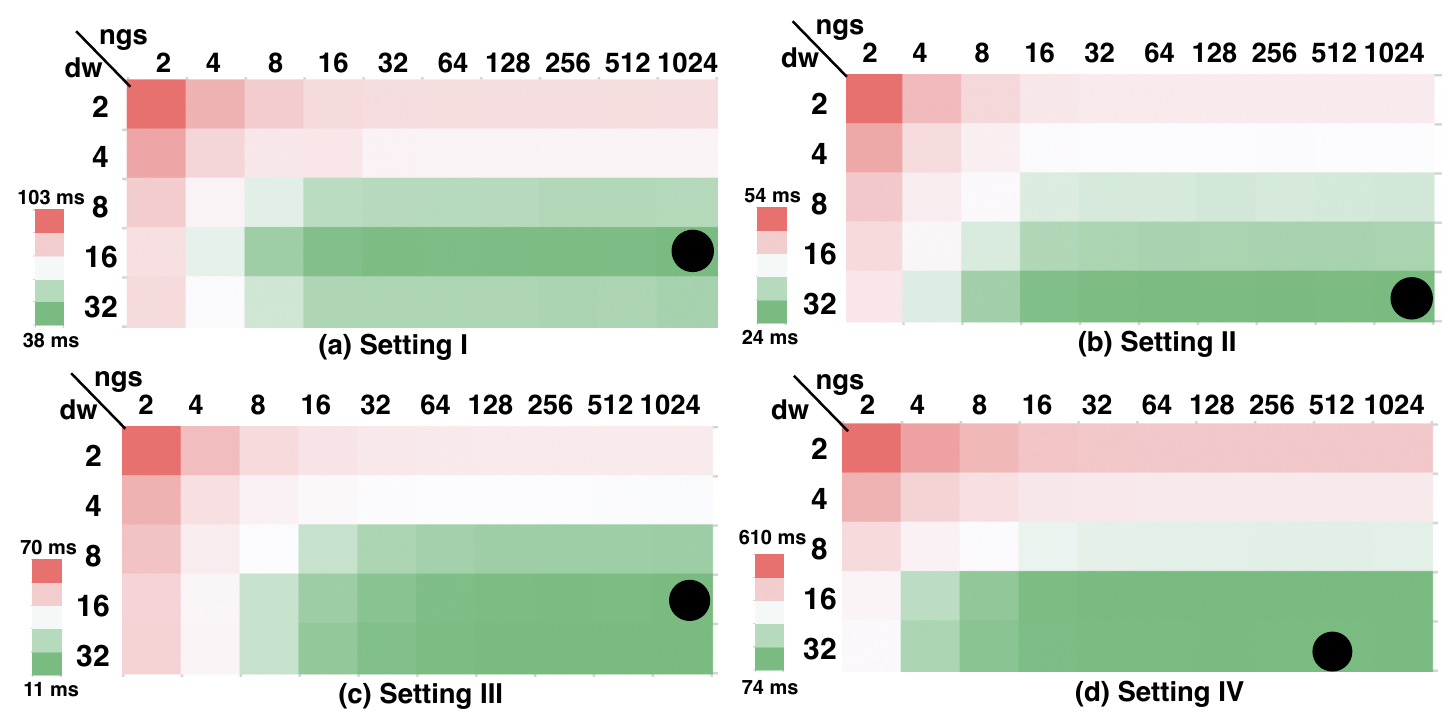}
    \vspace{-10pt}
    \caption{Parameter Selection for Four Settings. Note that the solid-black dot indicates the parameter ($\mathit{dw}$ and $\mathit{ngs}$) selected by \Mname~\textbf{\code{Decider}} based on analytical modeling.}
    \vspace{-15pt}
    \label{fig: Parameter Selection for Four Settings}
\end{figure}

\textbf{Parameter selection:}
To show the effectiveness of our analytical modeling in kernel parameter selection, we consider four different settings: 
I: \textit{amazon0505} on GCN at P6000 GPU as our base setting; 
II:  \textit{amazon0505} GCN on V100 to demonstrate device adaptation; 
III: \textit{amazon0505} and \textit{soc-BlogCatalog} on P6000 to demonstrate adaptation to different datasets;
IV: \textit{amazon0505} on GIN at P6000 to demonstrate adaptation to a different GNN model architectures.
As shown in Figure~\ref{fig: Parameter Selection for Four Settings}, our parameter selection strategy can pinpoint the optimal low-latency design for the above four settings. 
This demonstrates the effectiveness of our analytical modeling in assisting parameter selection to optimize the performance of GNN computation.
\vspace{-10pt}
\section{Conclusion}
\vspace{-5pt}
In this work, we propose, \Mname, an adaptive and efficient runtime system for GNN acceleration on GPUs. 
Specifically, we explore the potential of GNN input-level information in guiding system-level optimizations. 
We further propose a set of GNN-tailored system-level optimizations (\textit{e.g.}, 2D workload management, and specialized memory optimizations) and incorporate them into our parameterized designs to improve performance and adaptability. 
Extensive experiments on a wide range of datasets and mainstream GNN models demonstrate the effectiveness of our design.
Overall, \Mname~provides users a handy tool to accelerate GNNs on GPUs systematically and comprehensively.

\vspace{-10pt}
\section{Acknowledgment}
\vspace{-5pt}
We would like to thank our shepherd, Petros Maniatis, and the
anonymous OSDI reviewers.
This work was supported in part by NSF 1925717. 
Use was made of computational facilities purchased with funds from the National Science Foundation (OAC-1925717) and administered by the Center for Scientific Computing (CSC). The CSC is supported by the California NanoSystems Institute and the Materials Research Science and Engineering Center (MRSEC; NSF DMR 1720256) at UC Santa Barbara.

\bibliographystyle{plain}
\bibliography{reference}

\clearpage
\appendix
\section{Artifact Appendix}

\subsection*{Abstract Summary}
\Mname~is an efficient and adaptive runtime system for GNN computing on GPUs. 
\Mname~consists of two parts. 
The first part is the host-side CPU program. It is responsible for dataset loading, runtime configuration generation, and invoking the GPU-side program. 
The second part is the device-side GPU program. It is responsible for the major computation of the GNN model on sparse neighbor-aggregation and dense node-update phase.
\Mname~improves the performance of GNN computing with its highly configurable and efficient 2D workload management and specialized memory design. Moreover, the runtime configuration generation on the host-side CPU program makes \Mname~more adaptive towards various kinds of input settings.

\subsection*{Artifact Checklist}

\begin{itemize}
\itemsep0em 
    \item \textbf{Link:} \url{github.com/YukeWang96/OSDI21_AE.git}.
    \item \textbf{Hardware}:
    \begin{itemize}
    \itemsep0em 
        \item Intel CPU x86\_64 with host memory >= 32GB. Tested on Intel Xeon Silver 4110 (8-core 16-thread) CPU with 64GB host memory.
        \item  NVIDIA GPU (arch>=$sm\_60$) with devcie memory >= 16GB. Tested on NVIDIA Quadro P6000 ($sm\_61$), Tesla V100 ($sm\_70$), and RTX3090 ($sm\_86$). Note that upon creating this artifact, we mainly evaluate our design on RTX3090. The execution time may be different across different devices but the overall trend of performance (speedup) is similar.
    \end{itemize}
    \item \textbf{OS \& Compiler}:
    Ubuntu 16.04+, GCC 7.5+, CMAKE 3.14+, CUDA 10.2+.
\end{itemize}

\subsection*{Environment Setup}
\noindent \textbf{Step-1: Setup the basic environment.} Two options:
\begin{itemize}
\itemsep0em 
    \item Setup the environment via Docker (\textbf{Recommended}).
    \item Setup via conda and pip.
\end{itemize}
Details of the above two options can be found in \texttt{README.md}.

\vspace{3pt}
\noindent \textbf{Step-2: Install GNNAdvisor Pytorch Binding.}
\begin{itemize}
\itemsep0em 
    \item Go to \texttt{GNNAdvisor/GNNConv}, then \texttt{python setup.py install} to install the \texttt{GNNAdvisor} modules.
    \item Go to \texttt{rabbit\_module/src}, then \texttt{python setup.py install} to install the rabbit reordering modules.
\end{itemize}

\noindent \textbf{Step-3: Download the graph datasets.} 
Our preprocessed graph datasets in \texttt{.npy} format can be downloaded via this link~\footnote{https://bit.ly/3ys86a5} (filename: \texttt{osdi-ae-graphs.tar.gz}).
Unzip the graph datasets \texttt{tar -zxvf osdi-ae-graphs.tar.gz} at the project root directory.
Note that node initial embedding is not included, and we generate an all 1s embedding matrix according to users input dimension parameter at the runtime for just performance evaluation.

\subsection*{Experiments}

\begin{itemize}
\itemsep0em 
    \item Running DGL baseline on GNN training (Figure 9).
    \begin{itemize}
    \itemsep0em 
        \item Go to \texttt{dgl\_baseline/} directory.
        \item \texttt{./0\_run\_gcn.sh} and \texttt{./0\_run\_gin.sh} to run DGL and generate \texttt{.csv} result for GCN and GIN.
    \end{itemize}
    \item Running PyG baseline on GNN training (Figure 10).
    \begin{itemize}
    \itemsep0em 
        \item Go to \texttt{pyg\_baseline/} directory.
        \item \texttt{./0\_run\_gcn.sh} and \texttt{./0\_run\_gin.sh} to run PyG and generate \texttt{.csv} result for GCN and GIN.
    \end{itemize}
    \item Running Gunrock for single SpMM (neighbor aggregation) kernel.
    \begin{itemize}
    \itemsep0em 
        \item Go to \texttt{Gunrock/} call \texttt{./build\_spmm.sh}. 
        \item \texttt{./0\_bench\_Gunrock.py} for profile \texttt{spmm}.
    \end{itemize}
    \item Running GNNAdvisor (Figure 9 and 10).
    \begin{itemize}
    \itemsep0em 
        \item Go to \texttt{GNNAdvisor/} directory.
        \item \texttt{./0\_run\_gcn.sh} and \texttt{./0\_run\_gin.sh} to run GNNAdvisor and generate \texttt{.csv} for GCN/GIN.
    \end{itemize}
    \item Running some additional studies (Figure 11(a,b,c), and 12(a)). Detailed commands of running all these studies can be found in \texttt{README.md}.
\end{itemize}

Note that accuracy evaluation are omitted for all implementations and each sparse kernels are tested via the unitest.py
We focus on the training evaluation of the GNNs, and the reported time per epoch only includes the GNN model forward and backward computation, excluding the data loading and some preprocessing.
Since the paper draft submission and the creation of this artifact, DGL has update several of its kernel library (from v0.52 to v0.60). In this comparion we focus on the latest DGL version (v0.60).
Based on our profiling on RTX3090 and Quadro P6000, our design would show minor speedup on the simple GCN model (2-layer and 16 hidden dimension), but show more evident speedup on more complicated GIN model (5-layer and 64 hidden dimension), which can still demonstrate the effectiveness of our optimizations.
Our observation is that on small Type I graphs, our frameworks achieve significant speedup for both GCN and GIN model on RTX3090 and Quadro P6000. On larger Type II and Type III datasets, our GIN model implementation would show more evident speedups.

\end{document}